# MSRANetV2: An Explainable Deep Learning Architecture for Multi-class Classification of Colorectal Histopathological Images

**Ovi Sarkar[1], Md Shafiuzzaman[1], Md. Faysal Ahamed[1], Golam Mahmud[1], Muhammad E. H. Chowdhury[2 *]**

[1]Department of Electrical & Computer Engineering, Rajshahi University of Engineering & Technology, Rajshahi-6204, Bangladesh. (ovisarkareceian@gmail.com, shafiuzzaman.ruet@gmail.com, faysalahamedjishan@gmail.com, mahmud.ece.ruet@gmail.com)
[2]Department of Electrical Engineering, Qatar University, Doha 2713, Qatar (mchowdhury@qu.edu.qa)

*Corresponding author:
Muhammad E. H. Chowdhury
Department of Electrical Engineering,
Qatar University, Doha 2713, Qatar
(mchowdhury@qu.edu.qa)

**Declaration of Competing Interest**
The authors affirm that they do not possess any identifiable conflicting financial interests or personal affiliations that might have influenced the findings presented in this paper.

**Institutional Review Board Statement:** Not applicable

**Informed Consent Statement:** Not applicable

**Data Availability**
The data supporting this study's findings are available from the corresponding author upon reasonable request.

**Funding**
The authors declare that no funding was obtained for this research.

**Authors' Contribution**
Ovi Sarkar: Conceptualization, Methodology, Software, Investigation, Writing – Original Draft.
Md Shafiuzzaman: Data Curation, Formal Analysis, Validation, Writing – Original Draft.
Md. Faysal Ahamed: Resources, Visualization, Software, Writing – Original Draft.
Golam Mahmud: Visualization, Software, Writing – Original Draft.
Muhammad E. H. Chowdhury: Conceptualization, Supervision, Writing – Review & Editing, Funding Acquisition.
All authors have read and approved the final manuscript.

**Acknowledgments**
The open-access publication of the article is supported by Qatar National Library (QNL).

# MSRANetV2: An Explainable Deep Learning Architecture for Multi-class Classification of Colorectal Histopathological Images

***Abstract***

Colorectal cancer (CRC) is a leading worldwide cause of cancer-related mortality, and the role of prompt precise detection is of paramount interest in improving patient outcomes. Conventional diagnostic methods such as colonoscopy and histological examination routinely exhibit subjectivity, are extremely time-consuming, and are susceptible to variation. Through the development of digital pathology, deep learning algorithms have become a powerful approach in enhancing diagnostic precision and efficiency. In our work, we proposed a convolutional neural network architecture named MSRANetV2, specially optimized for the classification of colorectal tissue images. The model employs a ResNet50V2 backbone, extended with residual attention mechanisms and squeeze-and-excitation (SE) blocks, to extract deep semantic and fine-grained spatial features. In comparison with existing architectures, MSRANetV2 presents a multi-scale residual attention fusion strategy with squeeze-and-excitation (SE) blocks to dynamically recalibrate channel-wise features and emphasize informative spatial regions, leading to enhanced representational richness and class separability. These improvements address common limitations in baseline convolutional neural networks (CNNs), such as vanishing gradient issues and weak localization within complex histological textures. We evaluated our model on a five-fold stratified cross-validation strategy on two publicly available datasets: CRC-VAL-HE-7K and NCT-CRC-HE-100K. The proposed model achieved remarkable average precision, recall, F1-score, AUC, and test accuracy of 0.9884 ± 0.0151, 0.9900 ± 0.0151, 0.9900 ± 0.0145, 0.9999 ± 0.00006, and 0.9905 ± 0.0025, respectively, on the 7K dataset. Similarly, on the 100K dataset, it reached 0.9904 ± 0.0091, 0.9900 ± 0.0071, 0.9900 ± 0.0071, 0.9997 ± 0.00016, and 0.9902 ± 0.0006. Additionally, Grad-CAM visualizations were incorporated to enhance model interpretability by localizing tissue areas that are medically relevant. These findings validate that MSRANetV2 is a reliable, interpretable, and high-performing architectural model for classifying CRC tissues.

## 1. Introduction

Colorectal cancer (CRC) is one of the most severe health challenges globally. According to the American Cancer Society, there will be over 1.55 million new cases of cancer and 53,000 deaths from colorectal cancer in the US [1]. Colorectal cancer is prevalent in the large intestine due to uncontrollable cell growth caused by gene mutation. Benign tumors or polyps in the colon or rectum can evolve into malignant tumors and are among the leading precursors to CRC [2].

Medical professionals have long been confronted with immense difficulty in the accurate diagnosis of colorectal cancer (CRC). Although conventional diagnostic methods, including colonoscopy and fecal occult blood tests, are widely used, they are not without their limitations. The tests tend to exhibit low sensitivity, risk of complications and restricted reliability because of anatomical variations specific to individual patients [3]. Hence, the demand for minimally invasive and precise diagnostic methods has greatly increased. Recent developments in machine learning and digital pathology have introduced new opportunities for improving and automating the diagnosis process [4]. Digital pathology allows high-throughput analysis of whole-slide images and hence facilitates close examination and feature extraction. Traditional machine learning methods have made a significant contribution in this area by automating certain aspects of histological analysis; they are typically based on manually crafted features, which can be subjective, time-consuming, and error-prone.

Yet, most recent deep learning architectures suggested for the task of CRC classification suffer from serious limitations. Most architectures either capture superficial spatial features or over-emphasize deeper semantic information, resulting in imbalanced multi-scale representation. Furthermore, most models depend on large-scale networks with high computational costs, making them clinically impractical to deploy. A critical limitation is also the absence of model explainability because black-box predictions inhibit clinical trust and decision-making. They require a standard parameterized model but accurate architecture that integrates spatial and semantic cues effectively, without compromising on interpretability through explainable AI.

To solve such difficulties, deep learning has become a promising option. Deep learning can learn image features from raw image data by itself without any human help [5]. Deep CNNs have been very successful in image classification of medical images and even outperform humans sometimes [6]. Because of such advancements, this research suggests a strong deep learning framework for the specific task of classifying colorectal tissue.

Recent advances in convolutional neural networks (CNNs) have significantly influenced the field of medical image analysis, allowing accurate detection, classification, and segmentation of a variety of diseases in radiology and pathology. These models exhibit superior performance in recognizing subtle visual patterns in high-dimensional medical data. Yet, one of the biggest challenges is the opaqueness of deep learning models, which is often beset by a lack of explanation. In order to tackle this challenge, explainable artificial intelligence (XAI) techniques have been widely adopted to ensure interpretability and foster trust in clinical environments. XAI models, including Grad-CAM, provide qualitative insights into model predictions by highlighting the most significant regions in medical images. The combination of convolutional neural networks (CNNs) with interpretability methods has yielded promising results in numerous studies [7–11].

Our research aims to accomplish two primary objectives: one, to create a deep learning architecture that is capable of appropriately extracting and fusing multi-scale features to improve colorectal tissue classification accuracy; and two, to incorporate explainable artificial intelligence (XAI) in the model pipeline to improve clinical interpretability without compromising model performance. To achieve these goals, we presented a model architecture that integrates residual attention mechanisms and channel recalibration modules to augment feature discrimination. It aims to provide a model that not only competes with existing state-of-the-art methods but also provides visual justification for its predictions, hence making it applicable for real-world medical applications.

In this study, we introduced an innovative deep-learning framework called MSRANetV2 that tackles significant shortcomings in existing CRC diagnostic models. MSRANetV2 is developed on the ResNet50V2 foundation and integrates multi-scale residual attention with squeeze-and-excitation (SE) blocks to improve semantic and spatial feature representations. Utilizing attention-based fusion of shallow and deep features, our model guarantees enhanced accuracy in identifying tissue patterns. We conducted a thorough five-fold stratified cross-validation to assess robustness and generalization. In addition, our architecture is validated with two benchmark datasets—CRC-VAL-HE-7K and NCT-CRC-HE-100K—and is compared to various leading pre-trained models. Ultimately, to enhance model interpretability, we utilized Grad-CAM to visualize class-discriminative areas in histopathological images. The major contributions of our work are listed below:

1) We introduced the MSRANetV2 architecture, integrating residual attention from ResNet50V2, improved by squeeze-and-excitation (SE) modules. This design enables enhanced feature representation by utilizing multi-scale spatial information, essential for histopathological image classification.
2) The proposed MSRANetV2 employs channel alignment and upsampling to efficiently combine deep and shallow features, facilitating the effective integration of semantic and spatial information for enhanced classification of colorectal tissue.
3) A five-fold cross-validation method was utilized to guarantee a reliable and impartial assessment. The model underwent training and testing on various image splits, showcasing strong generalization ability on unfamiliar data.
4) The MSRANetV2 model outperformed several popular pre-trained architectures on the same dataset, achieving superior accuracy and F1 score across all folds.
5) To enhance interpretability, we incorporated Grad-CAM-based visual explanations, which provided class-discriminative localization maps, helping to validate the model's focus on medically relevant tissue regions.

## 2. Related Works

Over the past decade, the field of colorectal cancer detection using histopathological images has witnessed a significant evolution. Initially, colorectal cancers were detected manually by pathologists through visual inspection of tissue samples, a process that was time-consuming and prone to human error. In the early exploration of colorectal image classification, simple yet effective models began to gain attention. Kather [12] utilized several textual descriptors to address a multi-class issue involving tumor epithelium and simple stroma in a dataset of 5,000 histological images. He proposed various classification methods, including the k-nearest neighbors (k-NN), an SVM, decision tree with the RUSBoost technique, and training the classifiers, the use of 10-fold cross-validation. The findings revealed that the SVM method provided the highest performance, achieving an accuracy of 87.4% across eight classes.

Through the advancement of deep learning, researchers have consistently attempted to make the classification models more efficient and accurate. One of the earlier notable works was carried out by S. M. I. Uddin et al. [13], in which they proposed using ResNet50V2 for the classification of an eight-class histopathology dataset containing 5,000 images. Through the use of preprocessing techniques such as normalization, noise reduction, resizing, and augmentation, they achieved an impressive accuracy of 95%. Although they were successful, this research didn't focus on utilizing explainability methods, which would increase in importance for later models. C. Bhatt et al. in [14] employed VGG16 to categorize images from three distinct datasets into 9 and 8 categories. Their model, subjected to preprocessing methods such as color normalization and noise reduction, attained a strong accuracy of 97.6%, despite the absence of any augmentation techniques. This demonstrated that even without the complex transformations seen in later models, simple architectures could still deliver promising results, especially with proper preprocessing. In [15], Vinod Kumar D. et al. began exploring pre-trained CNN models, specifically evaluating architectures like EfficientNetB6, ResNet34, VGG-19, MobileNetV2, and ResNet50 for classifying 25,000 high-quality histopathology images. Among these models, MobileNetV2 and ResNet50 stood out, both achieving 99% accuracy, precision, recall, and F1 scores. However, one limitation was that this study did not address image segmentation, an essential task in more complex diagnostic processes.

As the field evolved, more advanced techniques began to emerge. One such breakthrough came from [16] T. Gurumoorthi et al., who introduced a self-attention-based CNN to perform binary classification on 10,000 colorectal histopathology images. By incorporating resizing, normalization, and basic augmentations (like flip and rotation), their model achieved a remarkable accuracy of 99.8%. This marked a significant leap, showing that enhancing traditional CNNs with self-attention mechanisms could greatly improve performance in cancer classification tasks. In [17], A. Merabet and colleagues introduced two hybrid models, InceptionV3-CNN and InceptionV3-ResNet50, for classifying a limited dataset of 2,500 colon cancer images. Their application of resizing, normalization, and a strong augmentation process (featuring rotation, shift, shear, zoom, and flip) led to remarkable accuracies of 99.27% and 99.20%, respectively. This emphasized the possibility of merging several strong models to attain enhanced accuracy, even with limited datasets. In [18], A. Kanadath et al. proposed the CViTS-Net model, a combination of traditional Convolutional Neural Networks (CNNs) and Vision Transformers (ViT). The model, which was improved by adding skip connections, aimed at understanding both local and global dependencies in histopathology images. When evaluated using a dataset of 6,160 images spanning four classes, their model recorded an accuracy of 96.06%. This was a first try at merging the strength of convolutional neural networks (CNNs) and transformers, an endeavor which would later become a significant research direction in the following years. In [19], Mahaveerakannan R and colleagues introduced a Swin Transformer featuring an altered final layer to categorize colorectal tumors as either benign or malignant. After applying a thorough preprocessing and augmentation process (including reading, resizing, denoising, segmentation, and morphological smoothing), their model reached an outstanding accuracy of 99.81%. This demonstrated the power of transformer-based models, particularly when optimized for the task at hand. Meanwhile, in [20], M. P. Young et al. took a different approach, leveraging intra-domain transfer learning and ensemble learning techniques to address the challenge of extracting comprehensive histopathological features. Their model, tested on several publicly available histopathology datasets, achieved top-tier performance, with accuracies of 99.78% on the GasHisSDB dataset, 85.69% on the Chaoyang dataset, and 99.17% on the CPTAC-CCRCC dataset. The use of ensemble learning, combined with robust transfer learning, set a new benchmark for colorectal image classification. In [21], S. Majumder et al. introduced the SAWL-Net, a lightweight model for classifying colorectal, breast, and lung cancers. This model uniquely combined similarity metrics (Pearson Correlation Coefficient, Spearman Rank Correlation, and Cosine Similarity) with a wave conversion approach, improving feature extraction across different histopathological datasets. Their model achieved an impressive accuracy of 99.90%, underscoring that even lightweight models could now compete with larger, more complex ones in terms of performance.

Although larger pre-trained models tend to exhibit outstanding performance in medical image analysis in general, normal pre-trained models as feature extractors tend to perform poorly, especially when working with datasets that have a very high number of classification classes. Such light models may struggle to effectively capture the complicated inter-class differences and finer patterns of medical images. As a result, the consistency of their classification and overall performance measures, including precision, recall, and F1-score, tend to be sacrificed. This emphasizes the need for new and well-designed network architectures capable of alleviating the drawbacks of such models and maintaining computational efficiency.

## 3. Methodology

Figure 1 illustrates the overall schematic workflow of the proposed colorectal tissue classification system using the MSRANetV2 architecture. Two publicly available histopathology image datasets—NCT-CRC-HE-7K and NCT-CRC-HE-100K [22]—were utilized. Each dataset was processed through a 5-fold cross-validation scheme, where images were split into 80% training, 10% validation, and 10% testing in each fold to ensure robust performance assessment. In the preprocessing stage, all images were resized to 224×224 and rescaled to the [0, 1] range to match the input format expected by the ResNet50V2 backbone. The proposed MSRANetV2 model incorporates a multi-scale residual attention fusion mechanism, leveraging intermediate feature maps from two key layers of the backbone network. This fusion helps in capturing both high-level semantic and low-level spatial cues. The classification head predicts one of the nine colorectal tissue categories. Additionally, Grad-CAM was incorporated to provide explainable AI insights, emphasizing class-specific areas within the image. Ultimately, the model's effectiveness was assessed using confusion matrix, classification reports, and ROC-AUC analysis, validating its reliability and interpretability.

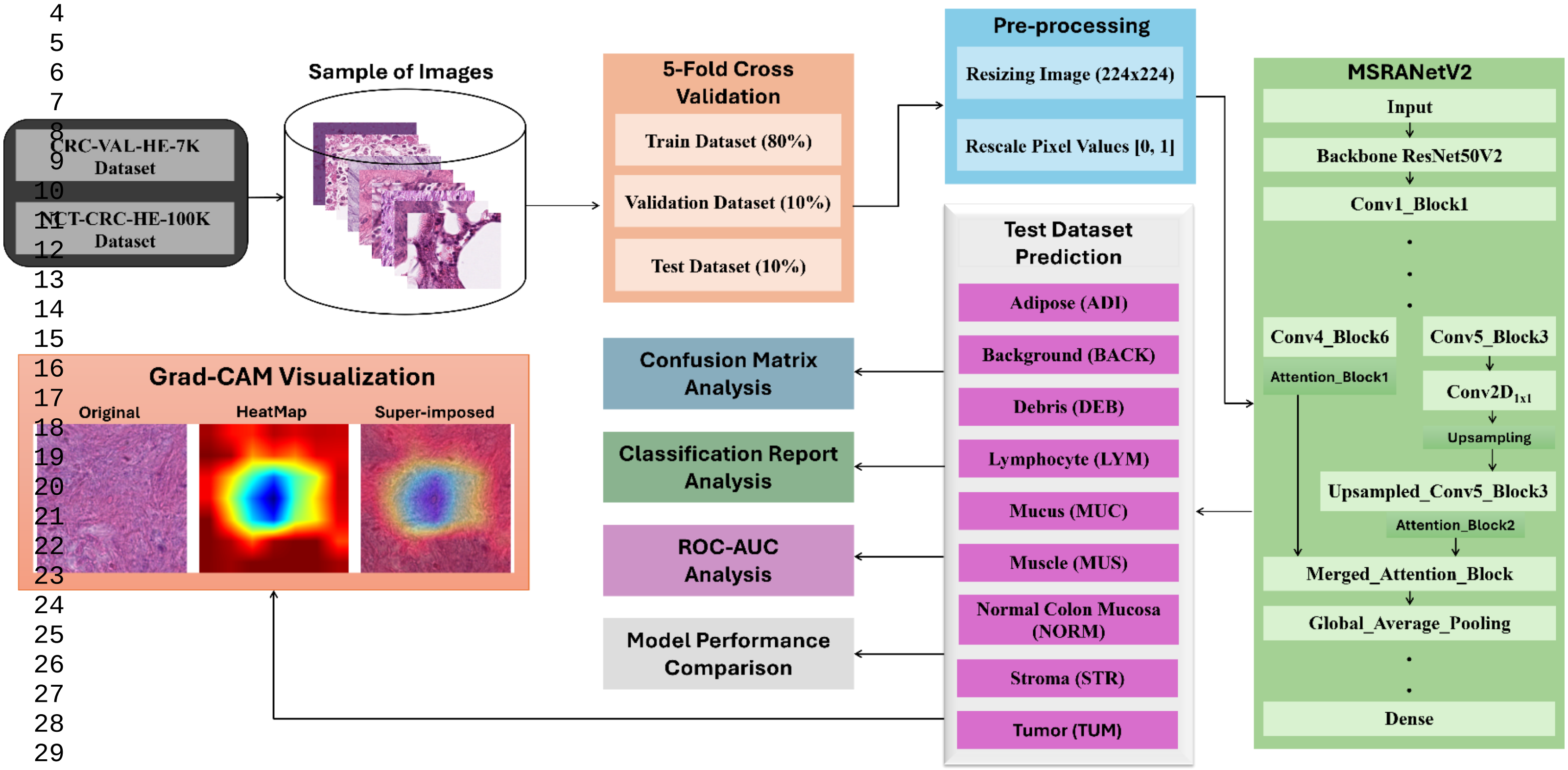

**Figure 1.** ***A schematic of the overall Multi-Scale Residual Attention Network V2 (MSRANetV2) system architecture.***

### 3.1 Dataset Description

In this research, two histopathology image datasets were employed to assess the efficiency of the suggested MSRANetV2 model: CRC-VAL-HE-7K and NCT-CRC-HE-100K. These datasets feature high-resolution images of hematoxylin and eosin (H&E) stained tissues classified into nine different colorectal tissue types. Both datasets are extensively utilized in computational pathology and provide varied, practical visual patterns crucial for effective model training and validation. The specific features of each dataset are outlined below.

#### 3.1.1 CRC-VAL-HE-7K

The CRC-VAL-HE-7K dataset was utilized in this research to evaluate the model's effectiveness in classifying colorectal cancer tissue images. It includes 7,180 colored images taken from tissue samples of 50 individuals with colorectal adenocarcinoma. A characteristic of this dataset is the exclusive patient distribution compared to its training dataset, aimed at enabling independent validation and reducing possible bias.

All images have undergone preprocessing and resizing to 224 × 224 pixels, with a spatial resolution of 0.5 microns per pixel (MPP). This standardization level allows for consistency throughout the dataset and is suitable for input into convolutional neural networks and other deep learning architectures. All tissue samples were ethically collected and provided by the NCT tissue bank. Images representative of each class is indicated in Fig. 2.

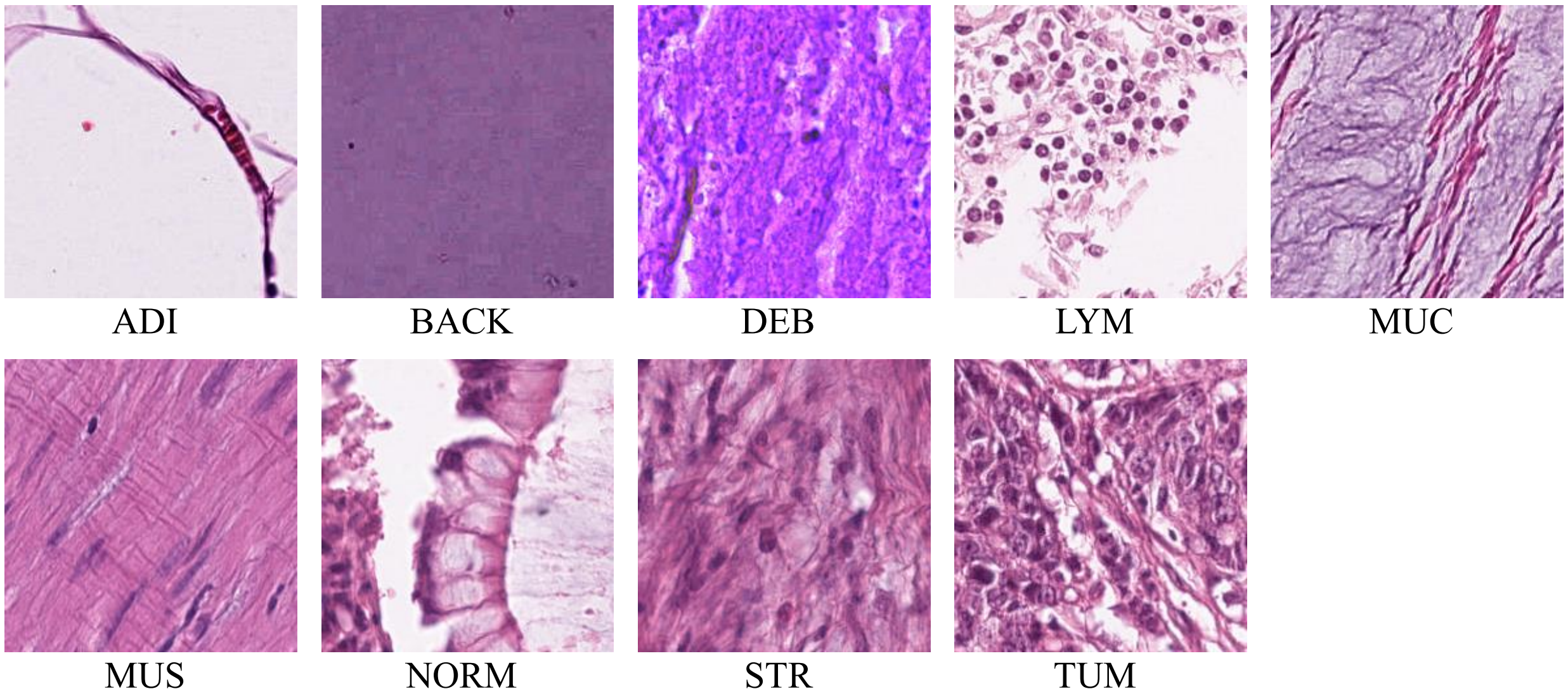

**Figure 2:** ***Visual samples of the nine tissue classes in the CRC-VAL-HE-7K dataset.***

The dataset is annotated into nine various histological classes, representing a wide range of tissue types that are found in colorectal histopathology. The classes are described below:

1. **ADI**: Adipose tissue mainly consists of adipocytes.
2. **BACK**: The background of a histological sample.
3. **DEB**: Debris is frequently used in histopathology and medical diagnoses.
4. **LYM**: Lymphocytes are the primary cell type found in the lymphatic system.
5. **MUC**: Mucus is produced by various tissues in the body, acting as a protective layer.
6. **MUS**: Tissue composed of smooth muscle.
7. **NORM**: Tissues from the normal colon mucosa.
8. **STR**: Stroma tissue associated with cancer.
9. **TUM**: Epithelial tissue from adenocarcinoma.

#### 3.1.2 NCT-CRC-HE-100K

The NCT-CRC-HE-100K dataset is a huge collection of histopathological images. It is designed to assist in the development and testing of machine-learning models that classify colorectal cancer tissues. In this study, 76,500 images were chosen carefully, ensuring a balanced class distribution across nine classes, with 8,500 samples per class. This balancing aids in a robust learning process and minimizes the risk of class imbalance while training.

All images in the dataset are derived from hematoxylin and eosin (H&E)-stained tissue samples collected from 86 patients diagnosed with colorectal adenocarcinoma. Each image was standardized to 224 × 224 pixels at a spatial resolution of 0.5 microns per pixel (MPP), ensuring uniformity suitable for deep learning pipelines. The data was collected, digitized, and ethically provided by the National Center for Tumor Diseases (NCT) and the University Medical Center Mannheim, with comprehensive ethical approvals in place. The class descriptions are the same as CRC-VAL-HE-7K dataset. Images representative of each class is indicated in Fig. 3.

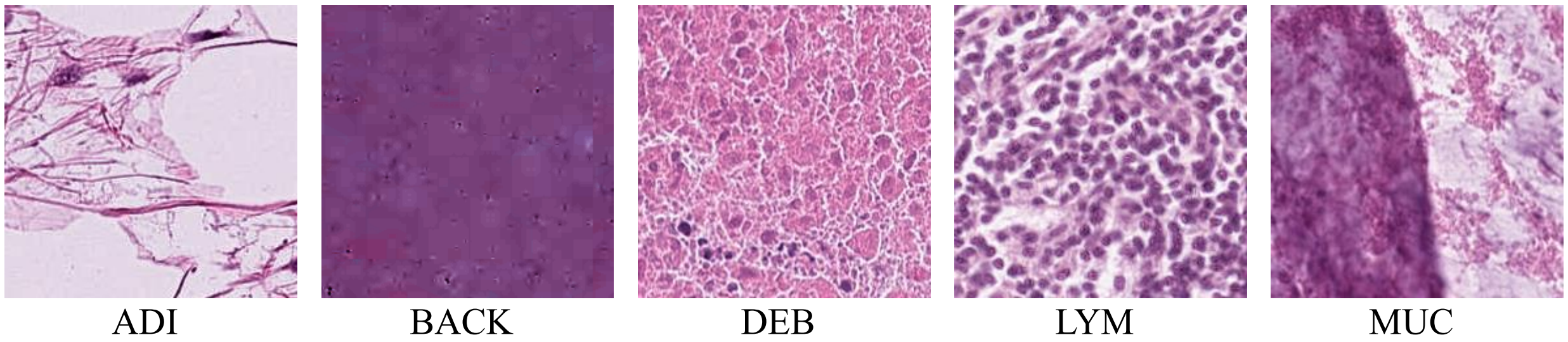

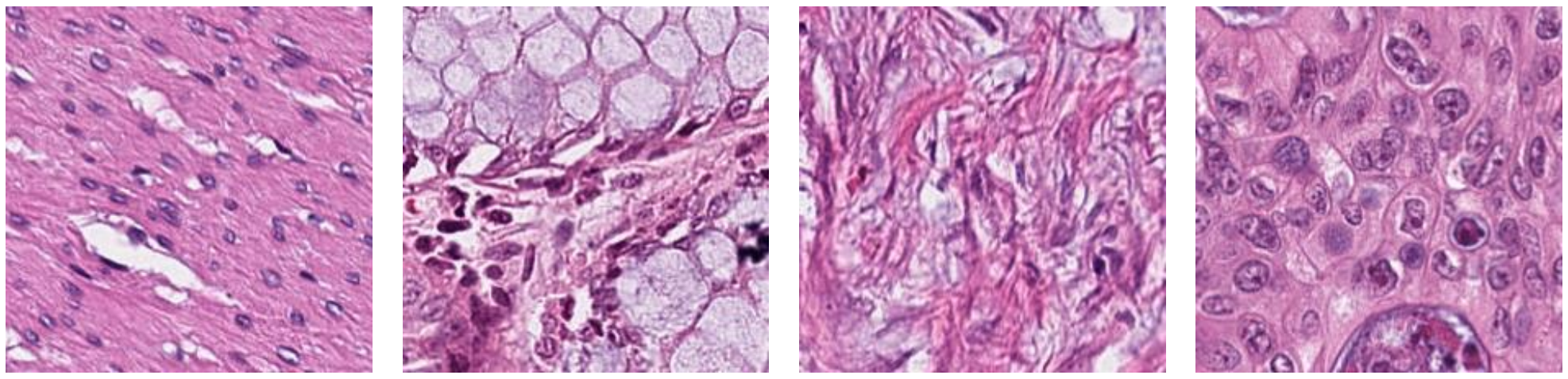

MUS NORM STR TUM

**Figure 3**: ***Visual samples of the nine tissue classes in the CRC-VAL-HE-100K dataset.***

### 3.2 Dataset Resizing

We experimented with multiple image resolutions of (128×128) and (224×224) to assess the impact of input size on model performance. Although resizing is commonly required for compatibility with pre-trained models like ResNet50V2 [23], both the CRC-VAL-HE-7K and NCT-CRC-HE-100K datasets natively contain images of size 224×224, eliminating the need for resizing. Empirically, we observed that using the original 224×224 resolution yielded the best classification performance, likely due to better preservation of histological details.

### 3.3 Normalization

To improve a model's efficiency and consistency, normalization is used to normalize each layer's inputs during training. Each pixel in this study is normalized by dividing it by 255, the maximum value that RGB photographs can have. Mathematical equation of this process can be written as follows:

$$x_{normalized} = \frac{x}{255}$$

Where, $x_{normalized}$ is the normalized pixel value, x is the original pixel value, 255 is the maximum possible value for a pixel in an RGB image. This process ensures that the values fall within the range of 0 to 1. This normalization helps the neural network learn more efficiently by standardizing the range of inputs to each layer.

### 3.4 Dataset Split

After resizing and normalization, the dataset is consistently split into training (80%), validation (10%), and testing (10%) sets using a stratified approach. This fixed ratio is maintained throughout all five folds. We apply 5-Fold Stratified Cross-Validation to ensure that each fold maintains the original class distribution, providing robust model evaluation and helping avoid issues of data imbalance across folds. Stratified splitting is particularly crucial for histopathology datasets, as class imbalance occurs frequently due to the uneven prevalence of tissue types or pathological conditions. In the absence of stratification, minority classes may be underrepresented in certain folds, resulting in biased training or unstable assessment. No other variations of splitting were utilized in this study.

### 3.5 Proposed Model Architecture (MSRANetV2)

The proposed MSRANetV2 (Multi-Scale Residual Attention Network Version 2) architecture in Figure 1 has two main components: a backbone and a convolution head for classification. The backbone is mainly a pre-trained network used for feature extraction in multi-class image classification. In our proposed architecture we have used ResNet50V2 as our backbone where we have removed the many convolution layers and used two specific bottom layer blocks for feature extraction. With the attention mechanism merging the two blocks and adding the convolution head of classification to use it as a multi-scale feature extractor with optimized filters, the model's ability largely enhanced its capability to extract the features for image classification. The main benefit of using transfer learning as a backbone is it leverages the pre-trained weights and helps to adapt specific characteristic elements capturing features from multiple scales. When the target task is classification, but the data is limited then it helps the model to perform better and achieve higher accuracy. This novel architecture involves a hybrid deep convolutional neural network with a residual and an attention-based mechanism, tailored specifically for multi-class image classification. The model is based on using the ResNet50V2 backbone with dual attention modules along with a feature fusion mechanism as well as a classification head. The architectural sequence occurs in the following order:

### 3.5.1 Input Layer

The MSRANetV2 architecture's input layer receives samples of colorectal images having a spatial resolution of 224×224 pixels and three-color channels (RGB), hence resulting in an input shape of (224, 224, 3). These images are histopathological images that are representative of different classes of colorectal tissue conditions. To maintain uniformity, standardization, and resizing are enabled within the dataset, thereby enhancing the feature extraction efficiently. The input tensor is directly fed into an adjusted ResNet50V2 backbone that serves as the basis for extensive multiscale learning and classification.

### 3.5.2 Feature Extraction using ResNet50V2 Backbone

The pre-trained model ResNet50V2 is employed as the feature extraction backbone in our proposed architecture. It is initialized with pre-trained weights on ImageNet with the removal of the last layers of classification. Instead of employing all the convolutional layers, we consciously extract two key stages of intermediate feature maps: conv4_block6_out (14×14×1024) and conv5_block3_out (7×7×2048). These layers are chosen because they strike a balance between spatial detail and semantic abstraction. The conv4_block6_out layer is at a shallower depth, hence retaining better spatial resolution and low-to mid-level features that are essential for preserving morphological and structural information in histopathology images. In contrast, the conv5_block3_out captures more abstract and class-specific high-level features. This fusion enables the model to conduct multi-scale feature fusion, effectively merging spatial granularity and semantic richness, which is especially beneficial in medical image analysis, where tissue architecture and fine-grained patterns are both essential.

To balance the dimensionality for feature fusion, the channel depth of conv5_block3_out from 2048 to 1024 was reduced using a 1×1 convolution. Then bilinear upsampling was applied to balance the spatial resolution and increase it from 7×7 to 14×14. The reason behind choosing bilinear upsampling over other methods is its computational efficiency and reduced risk of introducing checkerboard artifacts that are typically associated with transposed convolutions. Unlike nearest-neighbor interpolation, bilinear upsampling supports smoother transition and improved spatial continuity, which is essential while fusing attention-weighted features at different scales. Such a two-stage feature alignment assures a seamless fusion of shallow and deep features and thereby enhances classification performance on multi-class colorectal tissue images. The adoption of ResNet50V2 offers good gradient propagation and strong representational power through its deep residual structure.

### 3.5.3 Processing High-Level Features

The high-level features, once aligned with respect to size and depth, are passed through an attention mechanism that determines the most informative regions of the feature map. Using a squeeze-and-excitation strategy, the attention block computes global context through average pooling and reweight channels adaptively. The objective is to enhance the ability by emphasizing discriminative patterns and suppressing background noise irrelevant to the pattern. This step enriches the semantic depth of the high-level features and makes the model more sensitive to slight tissue structure differences. This step guarantees that the most informative high-level indicators are considered in the ultimate classification by concentrating the network's attention, thus improving overall precision and reliability.

### 3.5.4 Attention Mechanisms

To combine multi-scale semantic and spatial representations with efficiency, our proposed MSRANetV2 model employs a residual attention mechanism. This mechanism merges attention-enhanced elements extracted from two distinct stages of the ResNet50V2 backbone: conv4_block6_out and conv5_block3_out. Let the extracted feature maps be denoted as:

$F1 \in \mathbb{R}^{14\times14\times1024}$: output from conv4_block6_out

$F2 \in \mathbb{R}^{7\times7\times2048}$: output from conv5_block3_out

**Step 1: Dimensional Alignment and Upsampling**

To align dimensions for fusion, F2 is processed through a 1×1 convolution to decrease its channel depth and then upsampled to match F1:

$$F2' = \text{Upsample}\,(\text{Conv}_{1\times1}(F2)) \in \mathbb{R}^{14\times14\times1024}$$

**Step 2: Channel Attention using SE Block**

Each feature map (F1 and F2′) is passed through a shared channel attention block based on the Squeeze-and-Excitation (SE) mechanism.

**Squeeze**: Global Average Pooling is applied to each channel of the feature map to capture a summary statistic. This reduces the spatial dimensions (height and width) of the feature map into a single value per channel:

$z_c = (1 / H\times W) \times \Sigma_{i=1}^{H} \Sigma_{j=1}^{W} F_c (i, j)$, c=1, 2, ..., C

Here, $z_c$ is the squeezed scalar value (or summary statistic) for the c-th channel. It represents the global average of all pixel values in that channel, H is the height of the feature map (i.e., the number of pixels in vertical direction), W is the width of the feature map (i.e., the number of pixels in horizontal direction), and $F_c (i, j)$ is the activation value at position (i,j) in the c-th channel of the feature map.

**Excitation**: The squeezed vector **z** is then passed through two fully connected (dense) layers. The first layer reduces the number of channels to a lower-dimensional space (controlled by a reduction ratio **r**), and the second restores it back to the original dimension. A ReLU activation and a sigmoid function are applied to introduce non-linearity and normalize the output:

$s = \sigma (W_2 . \mathrm{ReLU}(W_1 . z))$

where $W_1 \in \mathbb{R}^{(C/r) \times C}$, $W_2 \in \mathbb{R}^{C \times (C/r)}$, r is the reduction ratio, and σ is the sigmoid activation function. This produces a vector of learned channel-wise weights s.

Here, s is the channel-wise attention vector (also called scaling factors), where each value $s_c \in [0,1]$ indicates the learned importance of the c-th channel. $W_1$ is the weight matrix of shape $\mathbb{R}^{(C/r) \times C}$. It reduces the channel dimension from C to c/r (where r is the reduction ratio). This is part of the first fully connected layer (used in the bottleneck structure of SE blocks). $W_2$ is the weight matrix of shape $\mathbb{R}^{C \times (C/r)}$. It restores the reduced vector back to the original channel size C through the second fully connected layer. z is the squeezed global context vector, obtained from global average pooling over the spatial dimensions. ReLU is the Rectified Linear Unit activation function used to introduce non-linearity after the first dense layer and Dot (.) denotes matrix multiplication (or dot product) between the weight matrices and the feature vectors.

A reduction ratio (r) of 16 was implemented for a good balance between model performance and complexity. A lower reduction ratio (e.g., 4 or 8) would make the excitation block more expressive by increasing its number of parameters and computations with the expense of more additional memory and longer training times. On the contrary, a larger value of r (i.e., 32) would simplify the network but might limit its ability to capture complex channel-wise relationships. Empirically, it was found that r = 16 provided the optimal trade-off, enabling effective attention weighting with minimal computational cost.

**Recalibration**: The original feature map F is multiplied (element-wise) with the learned weights s to emphasize the important channels and suppress the less relevant ones:

$F_{att} = s \odot F$

Here, $\odot$ denotes element-wise multiplication, and $F_{att}$ is the attention-refined feature map.

In histopathology images, Squeeze-and-Excitation (SE) blocks greatly enhance performance by enabling the model to concentrate on the most diagnostically useful features. Histopathology images often contain fine-grained cellular structures and intricate patterns that are dispersed across multiple channels. Through learning channel-wise dependencies, SE blocks emphasize the most informative channels and suppress the less relevant ones, thereby resulting in increased discriminative power. This mechanism is also highly effective in determining subtle inter-class differences in colorectal tissue types, where disparities in color and texture are very crucial.

This results in:

$F_1^{att} = SE(F1)$, $F_2^{att} = SE(F2')$

**Step 3: Residual Attention Fusion**

The two attention-refined feature maps are then combined using element-wise addition:

$F_{merged} = (F_1^{att} + F_2^{att}) \in \mathbb{R}^{14\times14\times1024}$

This residual addition keeps both the original and enhanced representations in the model so that it can preserve high-level semantic information along with fine-grained texture. This fusion strengthens the model's ability to distinguish subtle differences between colorectal histopathology images.

### 3.5.5 Classification Head

The merged feature map (14×14×1024) is fed to a Global Average Pooling 2D layer, giving a 1×1024 vector. Then it goes to a Dense layer with 512 units with ReLU activation, followed by Dropout (0.5) and Batch Normalization to help regularize and stabilize training. Ultimately, a Dense output layer with 9 units with softmax yields the class probabilities for colorectal tissue classification.

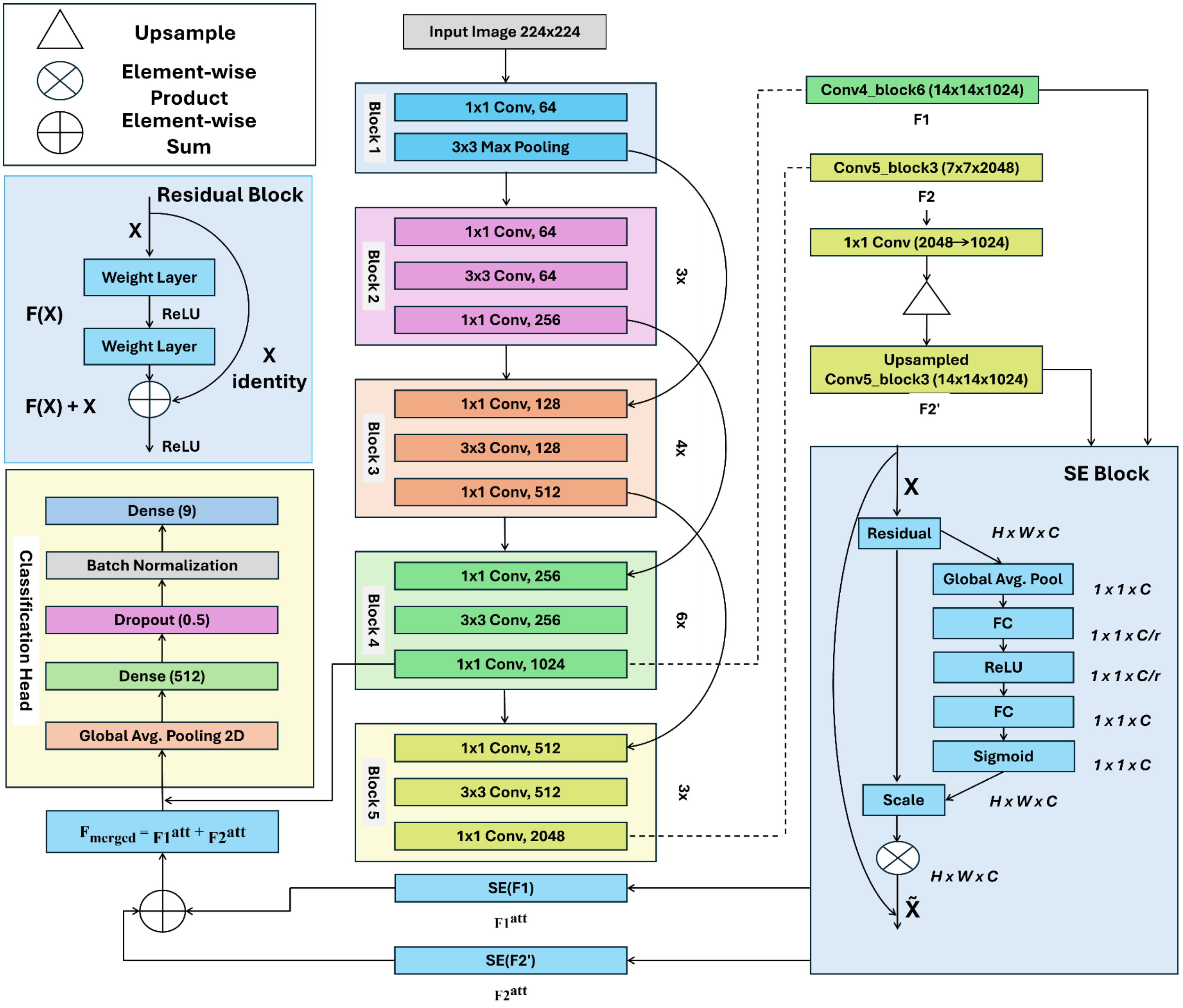

Figure 4: *Proposed architecture of MSRANetV2.*

**Algorithm:** Proposed MSRANetV2 Algorithm

1. **Input:** Histopathology images of colorectal tissues.
2. **Output Labels:** ADI, BACK, DEB, LYM, MUC, MUS, NORM, STR, TUM.
3. **Begin**
4. **Preprocessing:**
   i. Resize images: X*resized* = resize (X, 224×224×3)
   ii. Batch Normalization: X*norm* = Xresized/255
   iii. Prepare for 5-fold stratified split using StratifiedKFold
5. **For each fold k = 1 to 5, do:**
   **(a) Split:**
      i. Training set = 80%,
      ii. Validation set = 10%,
      iii. Test set = 10% (from fold-specific split)
   **(b) Attention-Based Feature Extraction:**
      i. Initialize base model: ResNet50V2 (input shape = 224×224×3)
      ii. Extract features:
         - F1=conv4_block6_out→14×14×1024
         - F2=conv5_block3_out→7×7×2048
      iii. Channel alignment: 1×1 Conv on F2: F2′ = Conv1x1(F2) → 7×7×1024
      iv. Upsample F2′: F2up→Bilinear Upsample(F2′) → 14×14×1024
   **(c) Squeeze-and-Excitation (SE) Attention Module:**
      i. For both F1 and F2up, apply SE block:
         - GlobalAvgPool → Dense (64) + ReLU → Dense (1024) + Sigmoid
         - Multiply channel-wise attention weights with feature maps
   **(d) Feature Fusion:**
      i. Fuse attention-enhanced maps: Ffused=F1⨁F2up→14×14×1024 (element-wise addition)
   **(e) Classification Head:**
      i. Global Average Pooling: G = GAP(Ffused)→1×1024
      ii. Dense layer: L1 = Dense (G, units = 512, activation = ReLU)
      iii. Batch Normalization: L2 = BatchNorm (L1)
      iv. Dropout: L3 = Dropout (L2, rate = 0.5)
      v. Output Layer: Ldense = Dense (L3, units = 9, activation = Softmax)
      vi. Compile: Mcompiled = Compile (Ldense, optimizer = Adam, learningrate=0.0001)
   **(f) Training Phase:**
      Train: Mtrained=train (Mcompiled, epochs = 15, batchsize = 16, validation = Xval)
   **(g) Testing Phase:**
      Predict labels: ytest = predict (Mtrained,Xtest)
6. **End**

### 3.6 Experimental Setup & Hyperparameter Settings

Table 1 presents the hardware and software configurations used for conducting all experiments in this study. The model was implemented using TensorFlow 2.15.0 and Keras 2.15.0, with GPU acceleration enabled through CUDA Toolkit 12.2 and cuDNN 8.9. A high-performance system equipped with an AMD Ryzen 9 7950X processor, 64 GB DDR5 RAM, and an NVIDIA GeForce RTX 4090 GPU with 24 GB GDDR6X memory ensured smooth training of deep learning models. All development and execution were performed in a Python 3.11 environment using Visual Studio Code on Windows 11 Pro. Table 1 provides a concise summary device configuration.

*Table 1: Hardware and Software configurations.*

| Name | Settings |
|---|---|
| Framework | TensorFlow 2.15.0, Keras 2.15.0, CUDA Toolkit 12.2, cuDNN 8.9 |
| RAM | 64 GB DDR5 |
| Processor | Model: AMD Ryzen 9 7950X, Clock Speed: 4.5 GHz up to 5.7 GHz, Cache: 80 MB, CPU Cores: 16, CPU Threads: 32 Socket: AM5 |
| GPU | NVIDIA® GeForce RTX™ 4090 Memory: 24 GB GDDR6X |
| Programming Language | Python 3.11 |
| Environment | Visual Studio Code 1.85.1 |
| Operating System | Windows 11 Pro |

*Table 2. Summary of selected hyperparameters and architectural choices used in the colorectal cancer tissue classification model.*

| Parameter Name | Combination Applied | Selected | Justification |
|---|---|---|---|
| Image Size | (128×128), (224×224) | (224×224) | Standard for ResNet-based models; balances accuracy and computational cost. |
| Stratify Split | Yes / No | Yes (category_encoded) | Ensures class distribution is balanced in train/test split. |
| Backbone Model | ResNet18, ResNet34, ResNet50, ResNet50V2 | ResNet50V2 | Offers better feature reuse and gradient flow compared to earlier versions. |
| Pretrained Weights | None, ImageNet | ImageNet | Facilitates transfer learning and faster convergence. |
| Attention Mechanism | None, Squeeze-and-Excitation (SE), CBAM, Custom | Custom (channel-wise scaling + SE) | Improves feature localization using channel-level weighting. |
| Pooling Layer | GlobalAveragePooling2D, Flatten | GlobalAveragePooling2D | Reduces dimensionality while preserving spatial features. |
| Dropout Rate | 0.3, 0.5, 0.6 | 0.5 | Balances regularization without underfitting; prevents overfitting. |
| Optimizer | Adam, SGD, RMSprop | Adam | Adaptive learning and widely effective for CNNs. |
| Learning Rate | 0.001, 0.0001, 0.00001 | 0.0001 | Provided stable convergence during trial runs. |
| Loss Function | Categorical Cross entropy, Focal Loss | Categorical Cross entropy | Appropriate for multi-class classification with softmax outputs. |
| Batch Size | 8,16, 32 | 16 | Balanced memory usage and gradient estimation quality. |

Hyperparameter optimization in deep learning is a non-trivial and computationally demanding task. In this study, we utilized a trial-and-error approach to determine suitable settings for colorectal cancer tissues detection. The trial-and-error approach was selected primarily because of the high computational costs of different hyperparameter optimization techniques like Bayesian optimization. Due to the depth and complexity of our architecture, associated with the use of large histopathology datasets, comprehensive systematic searches across various combinations of hyperparameters would have required considerable GPU time and memory resources. Trial-and-error allowed the model to be iteratively refined with informed intuition and empirical observations obtained from initial experiments, which was a feasible option in a time-constrained high-resource environment.

Table 2 provides a comprehensive summary of the hyperparameters, and architectural decisions adopted for colorectal tissue classification. Image size of 224×224 was selected, aligning with the ResNet50V2 input standard and offering a balance between performance and efficiency. Stratified splitting ensured a uniform class distribution across all folds. ResNet50V2 was chosen as the backbone due to its superior feature reuse and gradient propagation. Pretrained ImageNet weights were utilized for effective transfer learning and faster convergence. A custom attention mechanism combining channel-wise scaling and SE modules enhanced feature localization. Additional selections such as GlobalAveragePooling, dropout rate of 0.5, Adam optimizer, and a learning rate of 0.0001 were empirically validated for robust model training.

### 3.7 Classification Matrices and Loss Function

In this section, we present the experimental outcomes of our proposed colorectal cancer tissue classification framework using two different histological image datasets. For each dataset, we report the classification results obtained using cross-validation to assess the robustness and generalizability of our models. The performance is evaluated using standard metrics, and detailed analysis is provided in the subsequent subsections. Several performance metrics were utilized to evaluate the recognition capability of the proposed neural network architecture. These include the confusion matrix (CM), accuracy, precision, recall, F1-score, and the area under the receiver operating characteristic curve (AUC).

$$Accuracy = \frac{T_P + T_N}{T_P + T_N + F_P + F_N} \tag{1}$$

$$Precision = \frac{T_P}{T_P + F_P} \tag{2}$$

$$Recall = \frac{T_P}{T_N + F_P} \tag{3}$$

$$F1 - Score = \frac{2 * Precison * Recall}{Precision + Recall} \tag{4}$$

$$AUC = \frac{1}{2}\left(\frac{T_P}{T_P + F_N} + \frac{T_N}{T_N + F_P}\right) \tag{5}$$

Where $T_P =$ True positive, $T_N =$ True Negative, $F_P =$ False Positive, and $F_N =$ False negative.

The cross-entropy formula measures the performance of the model's predicted probability distribution against the actual class label, provided as an integer. The actual and predicted labels are compared with a view to reducing cross-entropy loss to the lowest possible. In deep learning applications that have more than one class, such as image classification, the sparse categorical cross-entropy loss is applied [24]. The formula is:

$$L_{ce} = -\sum_{i=1}^{n} y_r \times \log(y_p) \tag{6}$$

Where, $n$ denotes the class number, truth label is defined as $y_r$, and $y_p$ as the probability.

## 4 Experimental Results

This section provides an in-depth analysis of experimental results achieved by the proposed MSRANetV2 model. To ensure the reliability and generalizability of the model, extensive experiments were conducted on two benchmark colorectal cancer histopathology datasets: CRC-VAL-HE-7K and NCT-CRC-HE-100K. The CRC-VAL-HE-7K dataset provides a variety of histological patterns from 50 diverse patients, whereas NCT-CRC-HE-100K offers a large-scale image collection that is conducive to robust feature learning. Five-fold cross-validation was used to assess the model performance with stable training, validation, and test splits (80:10:10), thereby ensuring a fair and unbiased comparison between various runs. Throughout the experimentation process, important performance metrics like test accuracy, precision, recall, F1-score, and AUC score were calculated for both datasets. These metrics provide a better understanding of the model's classification ability, particularly in terms of dealing with various classes of colorectal tissue. Besides classification accuracy, explainable AI (XAI) techniques such as Grad-CAM were employed to highlight the model's interpretability and ability to focus on medically relevant regions. The interpretability results complemented the model predictions and aligned well with known pathological areas. All experimental results are tabulated and graphed for the sake of depicting patterns and model behavior. Results are presented systematically and discussed in the following subsections for both datasets.

### 4.1 Experimental Results: CRC-VAL-HE-7K Dataset

Table 3 shows the proposed MSRANetV2 architecture's five-fold cross-validation performance for colorectal cancer tissue classification using the CRC-VAL-HE-7K dataset across nine histological tissue classes. Each fold has a constant high classification ability, with macro and weighted averages of Precision, Recall, and F1-score at 0.99. Per-class performance remains consistently high, particularly for dominating classes such as Adipose (ADI), Background (BACK), and Mucus (MUC), which received perfect or near-perfect scores in the majority of folds. Classes with weaker support, such as Debris (DEB) and Stroma (STR), produced good results, demonstrating MSRANetV2's stability across class imbalances. The model achieved 99% overall accuracy across five folds, with a high mean AUC of 0.9998 ± 0.00008, demonstrating its discriminative capacity.

Figures 5 to 8 summarize the performance evaluation of the proposed MSRANetV2 model on the CRC-VAL-HE-7K dataset. Figure 5 illustrates accuracy trends, Figure 6 shows the loss progression, Figure 7 presents the ROC-AUC analysis, and Figure 8 displays the confusion matrix. In fold 1, the classifier achieves perfect accuracy for ADI, BACK, LYM, and MUS. NORM shows near-perfect performance with only one sample misclassified as LYM. Minor errors occur in TUM, MUC, STR, and DEB, each having one or two misclassifications, indicating overall strong and reliable classification across all classes. The classifier in fold 2 performed admirably overall, achieving perfect or near-perfect accuracy in the majority of classifications. BACK, DEB, LYM, and STR were all correctly categorized; however, MUS, MUC, and ADI each had one mistake, being confused with TUM and STR. NORM performed well but was misclassified once as LYM and once as TUM. TUM was the most perplexing target, with two inaccurate predictions from both MUC and STR. The classifier performed well overall, with perfect accuracy for BACK, DEB, LYM, and MUS, as these classes had predictions only along the main diagonal at fold 3. ADI and NORM also showed high accuracy with just one misclassification. A few misclassifications were noted, with TUM misclassified twice as ADI and DEB, STR misclassified three times, and MUC misclassified twice. The model performs well overall at fold 4, with most classes—ADI, BACK, DEB, LYM, and MUS—achieving perfect classification. Minor errors do occur in three classes, such as MUC has been categorized as TUM once, STR as MUS once, and TUM as LYM and NORM once. In the final fold, the classifier achieved perfect predictions for ADI, BACK, DEB, MUC, and NORM. Minor misclassifications were observed: STR was once predicted as TUM, TUM had two instances mislabeled as NORM, LYM was confused once with NORM, and MUS had one error where it was classified as STR.

Throughout the five folds, there were frequent misclassifications that were clear, especially in the classes Tumor (TUM), Stroma (STR), and Lymphocyte (LYM). Tumor (TUM) was consistently mistaken with MUC and STR, which could be due to the similar cell patterns and overlapping textures within regions of poorly differentiated tissue. Similarly, Stroma tissue (STR) was misclassified as MUS in most folds, which may be explained by their structural resemblance under fibrous regions when subjected to specific staining conditions. Lymphocytes (LYM) sometimes overlap with NORM, which is an indication of the challenge of separating sparse lymphocytic infiltration from normal mucosa. Such mistakes point out that while MSRANetV2 handles dominant classes with high accuracy, borderline tissue types that show morphological overlap still pose significant challenges. Future improvements might consider incorporating spatial context or tiles of higher resolution to alleviate such misclassifications.

**Table 3: Five-fold cross-validation performance metrics of MSRANetV2 for colorectal tissue classification across nine histological classes on the CRC-VAL-HE-7K dataset.**

| Fold Number | Classes | Precision | Recall | F1-score | Support | Accuracy | Avg. AUC |
|---|---|---|---|---|---|---|---|
| Fold 1 | Adipose (ADI) | 0.99 | 1.00 | 1.00 | 134 | 0.9916 | 0.9999 |
| | Background (BACK) | 1.00 | 1.00 | 1.00 | 84 | | |
| | Debris (DEB) | 1.00 | 0.97 | 0.99 | 34 | | |
| | Lymphocyte (LYM) | 0.98 | 1.00 | 0.99 | 63 | | |
| | Mucus (MUC) | 1.00 | 0.99 | 1.00 | 104 | | |
| | Muscle (MUS) | 0.98 | 1.00 | 0.99 | 59 | | |
| | Normal Colon Mucosa (NORM) | 0.99 | 0.99 | 0.99 | 75 | | |
| | Stroma (STR) | 1.00 | 0.95 | 0.98 | 42 | | |
| | Tumor Epithelium (TUM) | 0.98 | 0.99 | 0.99 | 123 | | |
| Fold 2 | Adipose (ADI) | 1.00 | 0.99 | 1.00 | 134 | 0.9875 | 0.9999 |
| | Background (BACK) | 1.00 | 1.00 | 1.00 | 84 | | |
| | Debris (DEB) | 1.00 | 1.00 | 1.00 | 34 | | |
| | Lymphocyte (LYM) | 0.98 | 1.00 | 0.99 | 63 | | |
| | Mucus (MUC) | 0.98 | 0.99 | 0.99 | 104 | | |
| | Muscle (MUS) | 1.00 | 0.98 | 0.99 | 59 | | |
| | Normal Colon Mucosa (NORM) | 1.00 | 0.97 | 0.99 | 74 | | |
| | Stroma (STR) | 0.93 | 1.00 | 0.97 | 43 | | |
| | Tumor Epithelium (TUM) | 0.98 | 0.97 | 0.97 | 123 | | |
| Fold 3 | Adipose (ADI) | 0.99 | 0.99 | 0.99 | 134 | 0.9875 | 0.9999 |
| | Background (BACK) | 1.00 | 1.00 | 1.00 | 85 | | |
| | Debris (DEB) | 0.97 | 1.00 | 0.99 | 34 | | |
| | Lymphocyte (LYM) | 0.98 | 1.00 | 0.99 | 63 | | |
| | Mucus (MUC) | 1.00 | 0.98 | 0.99 | 104 | | |
| | Muscle (MUS) | 0.97 | 1.00 | 0.98 | 59 | | |
| | Normal Colon Mucosa (NORM) | 0.99 | 0.99 | 0.99 | 74 | | |
| | Stroma (STR) | 0.93 | 0.93 | 0.93 | 42 | | |
| | Tumor Epithelium (TUM) | 1.00 | 0.98 | 0.99 | 123 | | |
| Fold 4 | Adipose (ADI) | 1.00 | 1.00 | 1.00 | 134 | 0.9930 | 0.9998 |
| | Background (BACK) | 1.00 | 1.00 | 1.00 | 85 | | |
| | Debris (DEB) | 1.00 | 1.00 | 1.00 | 34 | | |
| | Lymphocyte (LYM) | 0.98 | 1.00 | 0.99 | 63 | | |
| | Mucus (MUC) | 1.00 | 0.99 | 1.00 | 104 | | |
| | Muscle (MUS) | 0.98 | 1.00 | 0.99 | 59 | | |
| | Normal Colon Mucosa (NORM) | 0.99 | 0.99 | 0.99 | 74 | | |
| | Stroma (STR) | 0.98 | 0.98 | 0.98 | 42 | | |
| | Tumor Epithelium (TUM) | 0.99 | 0.98 | 0.99 | 123 | | |
| Fold 5 | Adipose (ADI) | 1.00 | 1.00 | 1.00 | 134 | 0.9930 | 1.0000 |
| | Background (BACK) | 1.00 | 1.00 | 1.00 | 85 | | |
| | Debris (DEB) | 1.00 | 1.00 | 1.00 | 33 | | |
| | Lymphocyte (LYM) | 1.00 | 0.98 | 0.99 | 64 | | |
| | Mucus (MUC) | 1.00 | 1.00 | 1.00 | 104 | | |
| | Muscle (MUS) | 1.00 | 0.98 | 0.99 | 59 | | |
| | Normal Colon Mucosa (NORM) | 0.96 | 1.00 | 0.98 | 74 | | |
| | Stroma (STR) | 0.98 | 0.98 | 0.98 | 42 | | |
| | Tumor Epithelium (TUM) | 0.99 | 0.98 | 0.99 | 123 | | |
| **Average (μ) ± SD ($\sigma$)** | | 0.9884 ± 0.0151 | 0.9900 ± 0.0151 | 0.9900 ± 0.0145 | – | 0.9905 ± 0.0025 | 0.9999 ± .00006 |

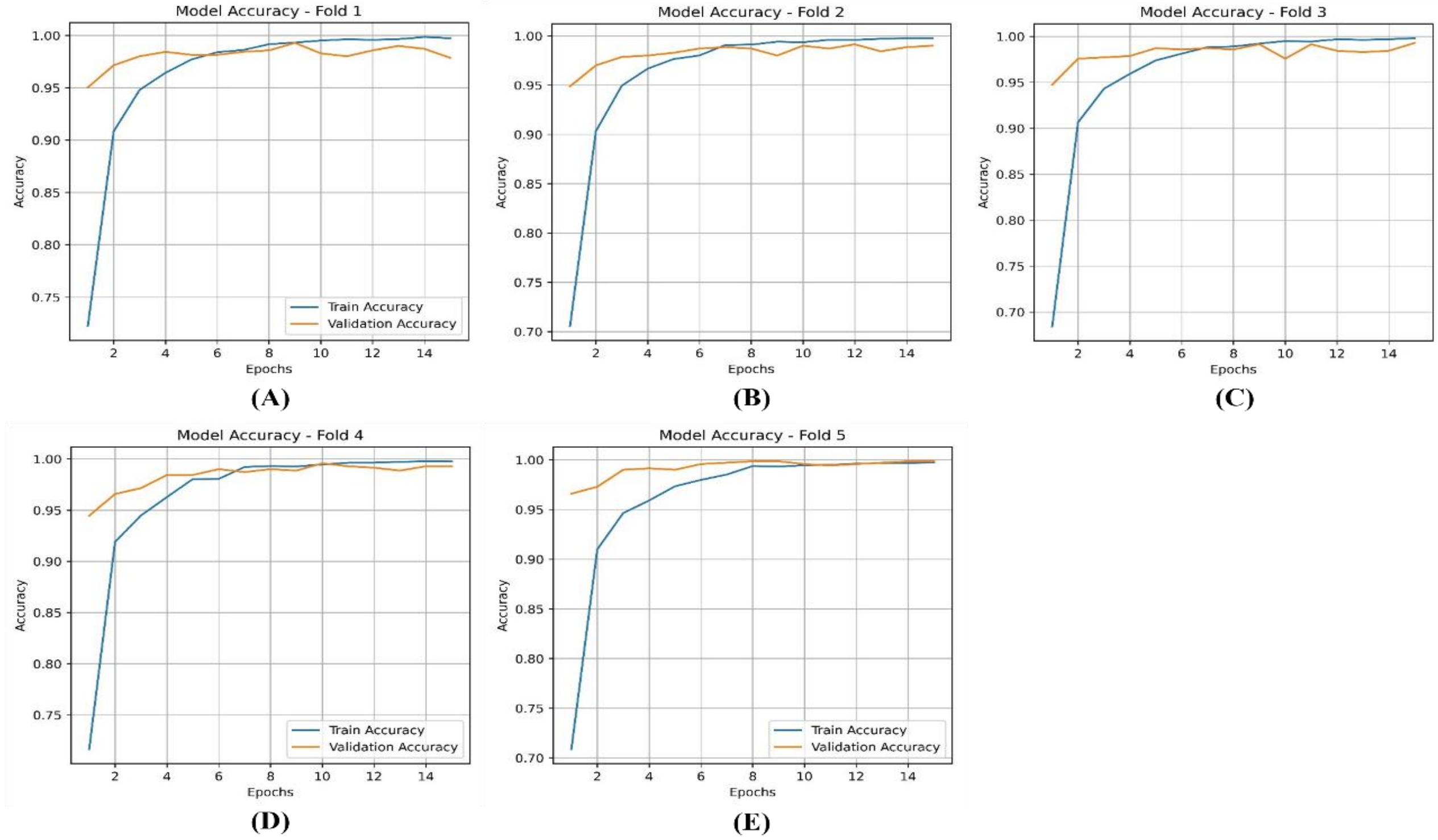

**Figure 5.** ***Training and validation accuracy of MSRANetV2 across five folds: (A) Fold 1, (B) Fold 2, (C) Fold 3, (D) Fold 4, and (E) Fold 5 on the CRC-VAL-HE-7K dataset.***

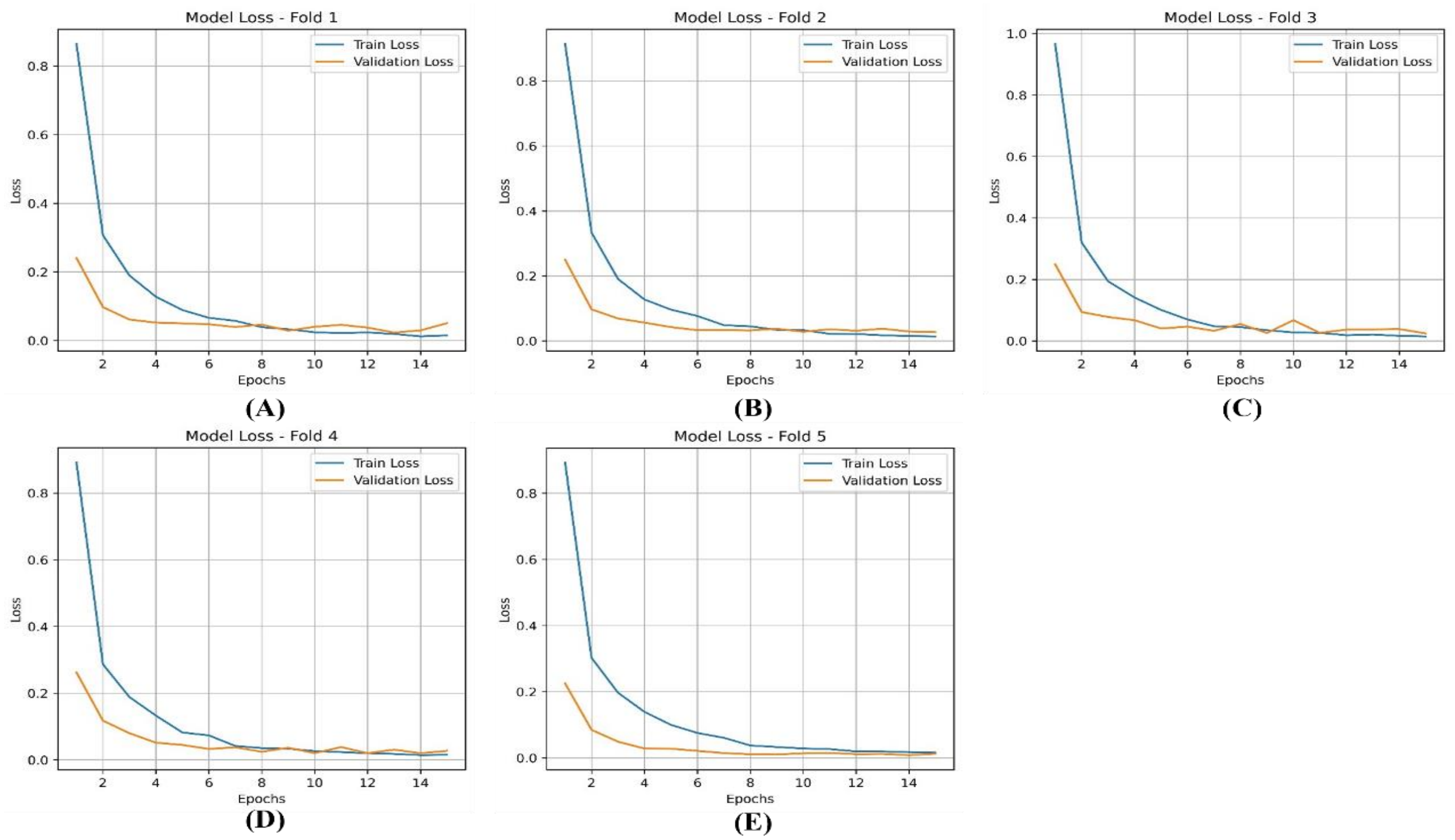

**Figure 6.** ***Training and validation loss of MSRANetV2 across five folds: (A) Fold 1, (B) Fold 2, (C) Fold 3, (D) Fold 4, and (E) Fold 5 on the CRC-VAL-HE-7K dataset.***

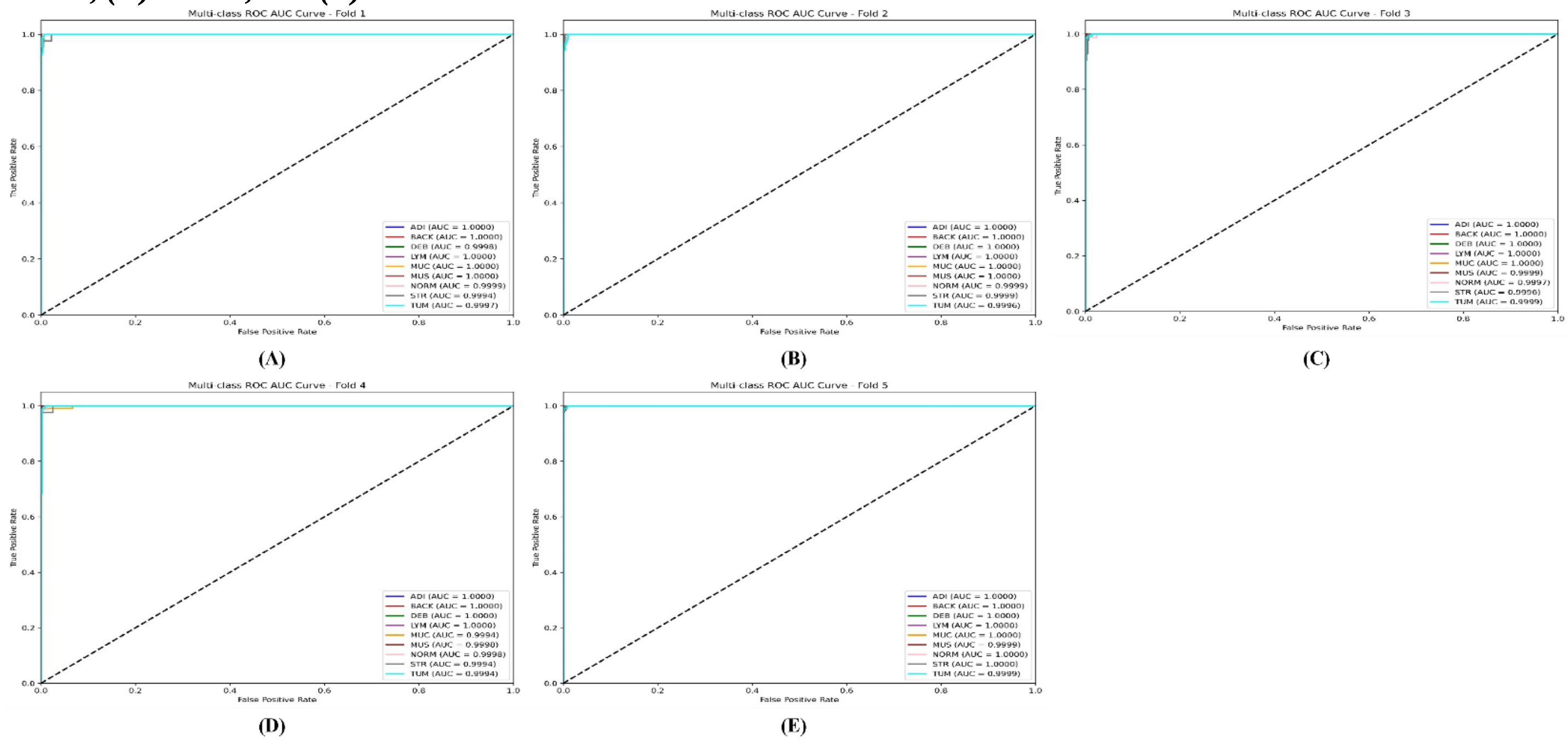

***Figure 7. ROC-AUC curves of MSRANetV2 for nine-class colorectal tissue classification across five folds: (A) Fold 1, (B) Fold 2, (C) Fold 3, (D) Fold 4, and (E) Fold 5 on the CRC-VAL-HE-7K dataset.***

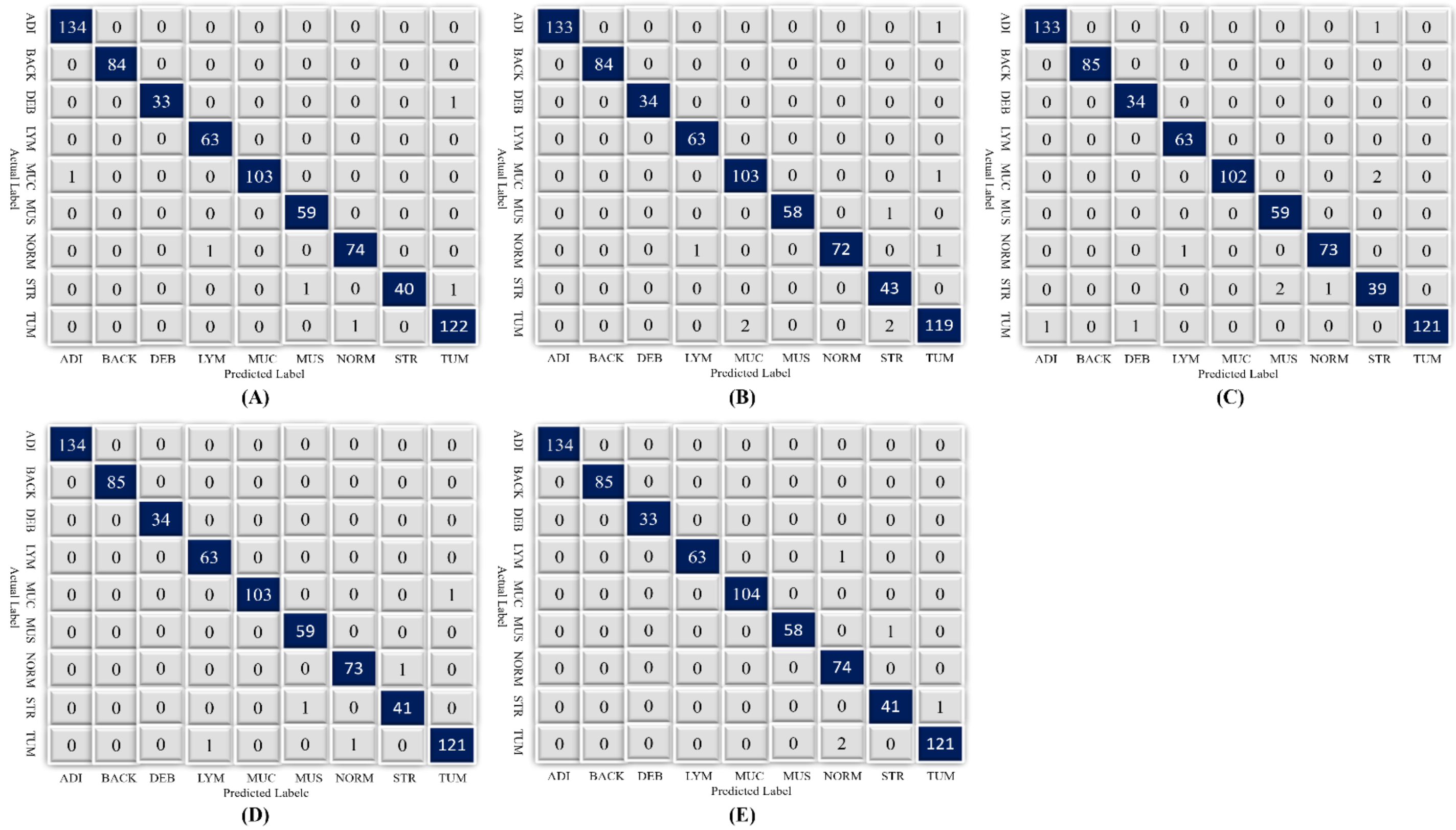

***Figure 8. Confusion matrices showing MSRANetV2's classification performance on five folds: (A) Fold 1, (B) Fold 2, (C) Fold 3, (D) Fold 4, and (E) Fold 5 on the CRC-VAL-HE-7K dataset.***

### 4.2 Experimental Results: NCT-CRC-HE-100K Dataset

Table 4 presents the detailed performance of MSRANetV2 for colorectal cancer tissue classification using five-fold cross-validation on the NCT-CRC-HE-100K dataset. The model was evaluated across nine histological classes: Adipose (ADI), Background (BACK), Debris (DEB), Lymphocyte (LYM), Mucus (MUC), Muscle (MUS), Normal Colon Mucosa (NORM), Stroma (STR), and Epithelium (TUM). For each fold, MSRANetV2 consistently achieved high scores in all evaluation metrics. ADI and BACK tissues attained perfect precision, recall, and F1-score (1.00) across all five folds, indicating excellent separability. Other classes, such as DEB, LYM, MUC, MUS, and NORM, also showed strong performance, with F1-scores ranging from 0.98 to 1.00. LYM received an F1-score of 0.99 in Fold 3 and 1.00 in Folds 1, 2, 4, and 5. Across all folds, MUC and MUS consistently received precision and recall scores of 0.98 or 0.99. The slightly reduced but still strong F1-scores of 0.97 to 0.98 for STR and TUM indicated very moderate classification difficulties. There were roughly 793–794 support samples in each class, guaranteeing a fair assessment. All folds had the same accuracy, recall, and F1-score macro and weighted averages of 0.99 ± 0.00. For every fold, the model consistently obtained an accuracy of 0.99. For Folds 1 through 5, the corresponding AUC scores were 0.9997, 0.9998, 0.9996, 0.9998, and 0.9993, yielding a mean AUC of 0.9964 ± 0.0001. In multi-class colorectal tissue categorization, these findings show MSRANetV2's stability, generalizability, and potent discriminative capacity.

Figures 1–4 show the overall performance of the proposed MSRANetV2 model on the NCT-CRC-HE-100K dataset. Figure 1 represents the training and validation accuracy, Figure 2 the loss curves, Figure 3 the ROC-AUC curve, and Figure 4 the confusion matrix. In figure 4, Fold 1 showed multiple misclassifications. MUC was commonly confused with STR three times, TUM three times, and NORM twice. On the other hand, STR was mis-predicted as DEB twice, MUC five times, and MUS seven times. In fold 2, the classifier does incredibly well on BACK and LYM, with no misclassifications. However, STR is the most prone to error, with ten instances of confusion with DEB and six with MUS. There are several

small mix-ups between NORM, TUM, or LYM. The classifier showed perfect accuracy for ADI and performed nearly flawlessly on LYM and BACK at fold 3. Most misclassifications occurred when MUS was mistaken for STR, and there was some confusion between STR and DEB. In Fold 4, the classifier performs admirably, with perfect accuracy on BACK and LYM, as well as good accuracy on ADI (792 out of 793) and MUC (790 out of 794). The most common misclassifications were DEB being forecasted as STR five times and STR as DEB three times. MUS was also wrongly identified as STR on six occasions. The classifier at fold 5 demonstrated outstanding accuracy for BACK and LYM, achieving perfect classification. ADI also showed strong performance with only twice misclassification. The most frequent errors occurred between MUS and STR, with 12 instances of misclassification, and between DEB and STR, with 5 errors. TUM was sometimes misclassified as NORM in 6 instances and as MUC in 4 instances.

Despite MSRANetV2's commendable performance across the various folds, specific histological classes exhibited persistent misclassification trends. The most prevalent confusion arose between MUS (muscle) and STR (stroma), likely attributable to overlapping fibrous textures and similarities in staining within these tissue types. This confusing pattern was consistently observed, with 6 to 12 misclassifications occurring across multiple folds. Furthermore, STR and DEB were frequently mistaken for one another, possibly due to the presence of fragmented stroma within debris regions, which complicates visual differentiation. Tumor tissue (TUM) was occasionally mistaken for either MUC or NORM, an effect that can be attributed to transitional morphological patterns between healthy mucosa and dysplastic or mucus-secreting tumor regions. The addition of higher-resolution tiles or contextual patch-level insight may help improve differentiation in future iterations of the model.

**Table 4: Five-fold cross-validation performance metrics of *MSRANetV2* for colorectal tissue classification across nine histological classes on the NCT-CRC-HE-100K dataset.**

| Fold Number | Classes | Precision | Recall | F1-score | Support | Accuracy | Avg. AUC |
|---|---|---|---|---|---|---|---|
| Fold 1 | Adipose (ADI) | 1.00 | 1.00 | 1.00 | 793 | 0.9901 | 0.9998 |
| | Background (BACK) | 1.00 | 1.00 | 1.00 | 793 | | |
| | Debris (DEB) | 0.99 | 0.99 | 0.99 | 793 | | |
| | Lymphocyte (LYM) | 1.00 | 1.00 | 1.00 | 794 | | |
| | Mucus (MUC) | 0.99 | 0.99 | 0.99 | 794 | | |
| | Muscle (MUS) | 0.99 | 0.99 | 0.99 | 794 | | |
| | Normal Colon Mucosa (NORM) | 0.99 | 0.99 | 0.99 | 793 | | |
| | Stroma (STR) | 0.97 | 0.98 | 0.97 | 794 | | |
| | Tumor Epithelium (TUM) | 0.99 | 0.98 | 0.98 | 793 | | |
| Fold 2 | Adipose (ADI) | 1.00 | 1.00 | 1.00 | 793 | 0.9898 | 0.9998 |
| | Background (BACK) | 1.00 | 1.00 | 1.00 | 793 | | |
| | Debris (DEB) | 0.99 | 0.99 | 0.99 | 794 | | |
| | Lymphocyte (LYM) | 0.99 | 1.00 | 1.00 | 793 | | |
| | Mucus (MUC) | 0.98 | 0.99 | 0.98 | 794 | | |
| | Muscle (MUS) | 0.99 | 0.99 | 0.99 | 794 | | |
| | Normal Colon Mucosa (NORM) | 0.99 | 0.98 | 0.98 | 793 | | |
| | Stroma (STR) | 0.98 | 0.97 | 0.98 | 793 | | |
| | Tumor Epithelium (TUM) | 0.99 | 0.98 | 0.98 | 794 | | |
| Fold 3 | Adipose (ADI) | 1.00 | 1.00 | 1.00 | 793 | 0.9894 | 0.9996 |
| | Background (BACK) | 1.00 | 1.00 | 1.00 | 793 | | |
| | Debris (DEB) | 0.99 | 0.99 | 0.99 | 794 | | |
| | Lymphocyte (LYM) | 0.99 | 1.00 | 0.99 | 794 | | |
| | Mucus (MUC) | 0.99 | 0.99 | 0.99 | 794 | | |
| | Muscle (MUS) | 0.99 | 0.98 | 0.99 | 793 | | |
| | Normal Colon Mucosa (NORM) | 0.99 | 0.98 | 0.99 | 793 | | |
| | Stroma (STR) | 0.97 | 0.99 | 0.98 | 794 | | |
| | Tumor Epithelium (TUM) | 0.98 | 0.99 | 0.98 | 793 | | |
| **Fold 4** | Adipose (ADI) | 1.00 | 1.00 | 1.00 | 793 | 0.9908 | 0.9998 |
| | Background (BACK) | 1.00 | 1.00 | 1.00 | 793 | | |
| | Debris (DEB) | 0.99 | 0.99 | 0.99 | 793 | | |
| | Lymphocyte (LYM) | 1.00 | 1.00 | 1.00 | 794 | | |
| | Mucus (MUC) | 0.98 | 0.99 | 0.99 | 794 | | |
| | Muscle (MUS) | 0.99 | 0.99 | 0.99 | 794 | | |
| | Normal Colon Mucosa (NORM) | 0.99 | 0.98 | 0.99 | 793 | | |
| | Stroma (STR) | 0.98 | 0.98 | 0.98 | 793 | | |
| | Tumor Epithelium (TUM) | 0.98 | 0.98 | 0.98 | 794 | | |
| **Fold 5** | Adipose (ADI) | 1.00 | 1.00 | 1.00 | 793 | 0.9910 | 0.9994 |
| | Background (BACK) | 1.00 | 1.00 | 1.00 | 793 | | |
| | Debris (DEB) | 1.00 | 0.99 | 0.99 | 793 | | |
| | Lymphocyte (LYM) | 1.00 | 1.00 | 1.00 | 794 | | |
| | Mucus (MUC) | 0.99 | 0.98 | 0.99 | 793 | | |
| | Muscle (MUS) | 0.99 | 0.98 | 0.99 | 793 | | |
| | Normal Colon Mucosa (NORM) | 0.99 | 0.99 | 0.99 | 794 | | |
| | Stroma (STR) | 0.97 | 0.99 | 0.98 | 793 | | |
| | Tumor Epithelium (TUM) | 0.98 | 0.98 | 0.98 | 794 | | |
| **Average (μ) ± SD (σ)** | | 0.9904 ± 0.0091 | 0.9900 ± 0.0071 | 0.9900 ± 0.0071 | – | 0.99022 ± 0.00060 | 0.99968 ± 0.00016 |

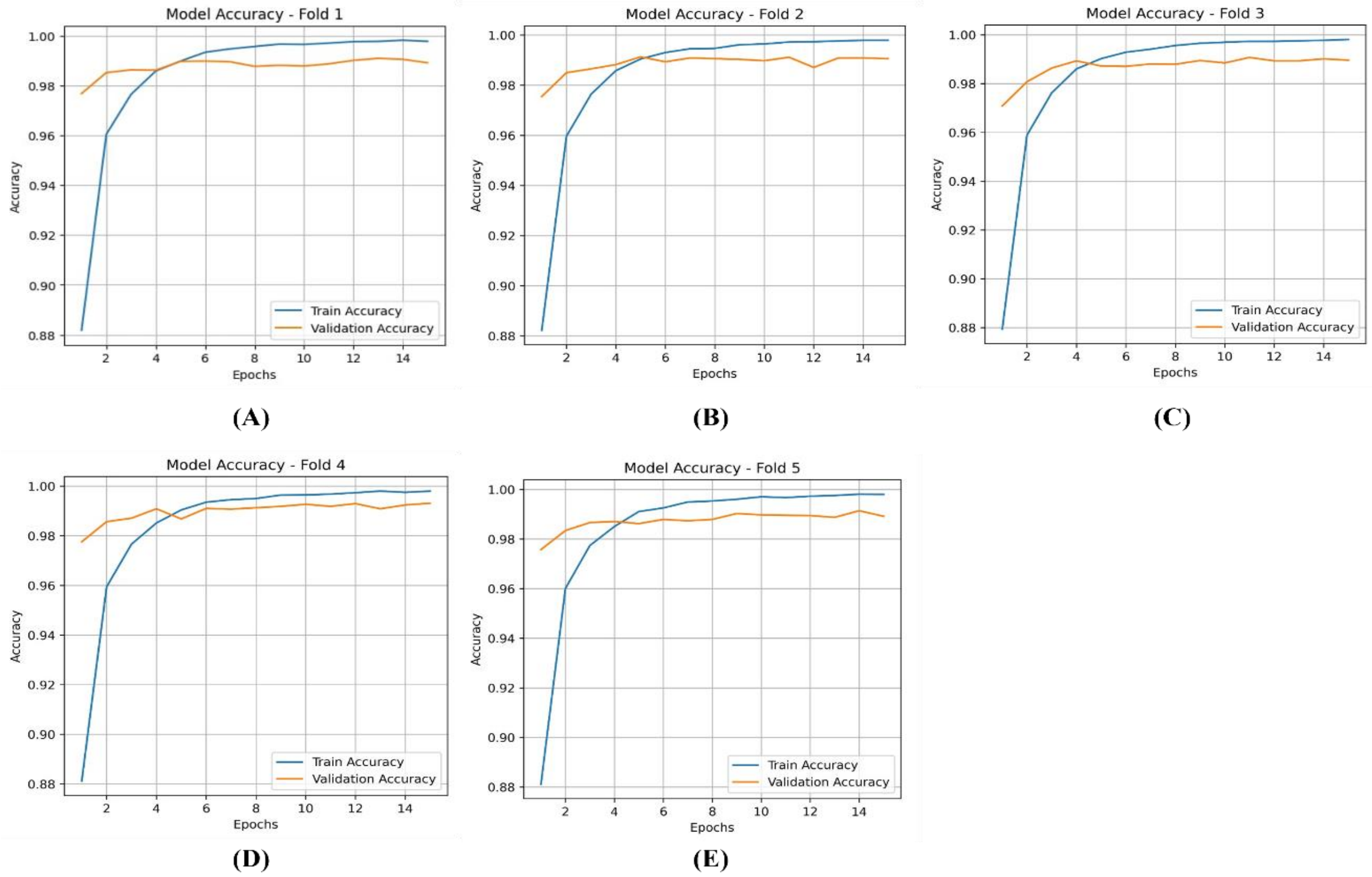

**Figure 9.** ***Training and validation accuracy of MSRANetV2 across five folds: (A) Fold 1, (B) Fold 2, (C) Fold 3, (D) Fold 4, and (E) Fold 5 on the NCT-CRC-HE-100K dataset.***

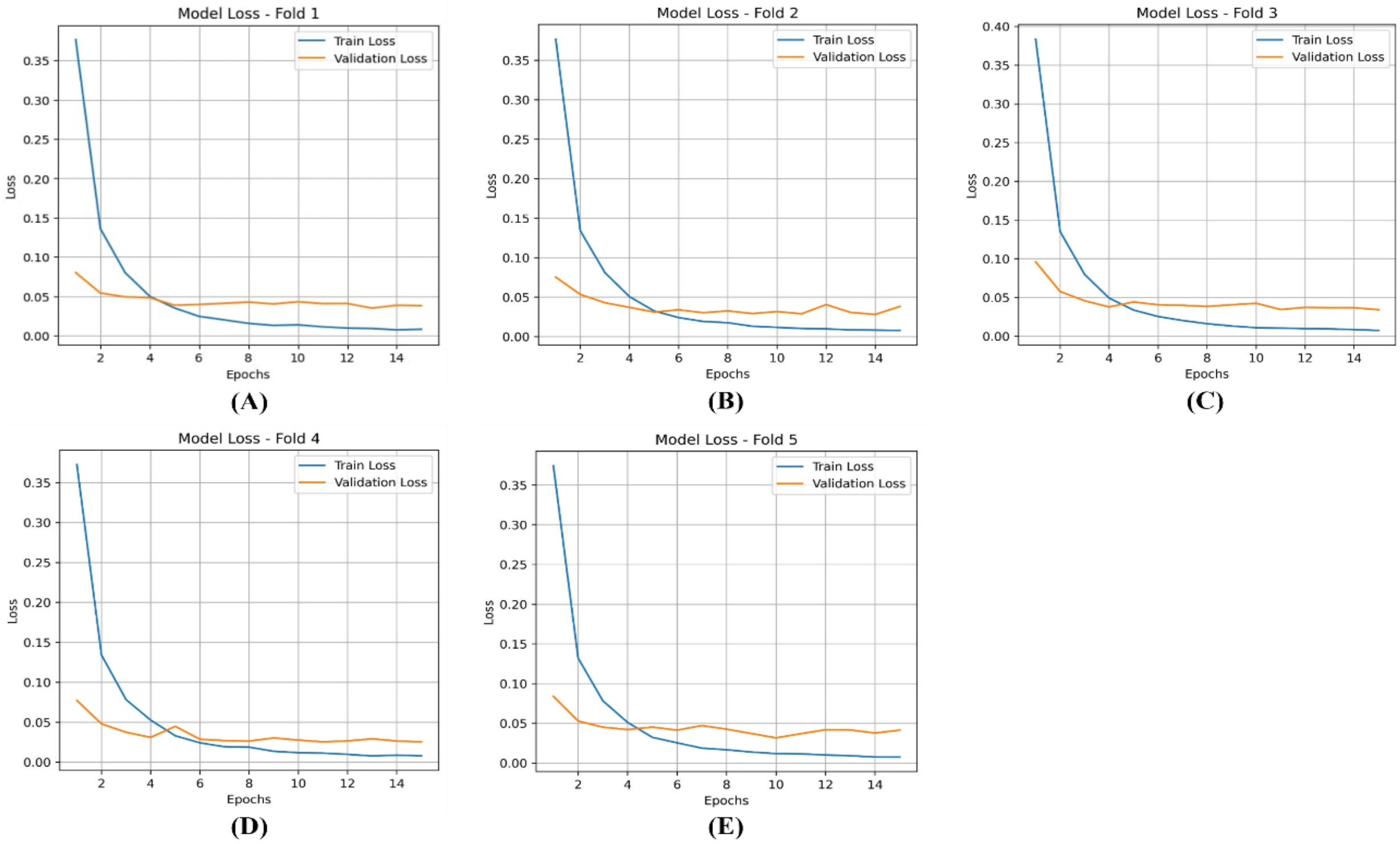

**Figure 10.** ***Training and validation loss of MSRANetV2 across five folds: (A) Fold 1, (B) Fold 2, (C) Fold 3, (D) Fold 4, and (E) Fold 5 on the NCT-CRC-HE-100K dataset.***

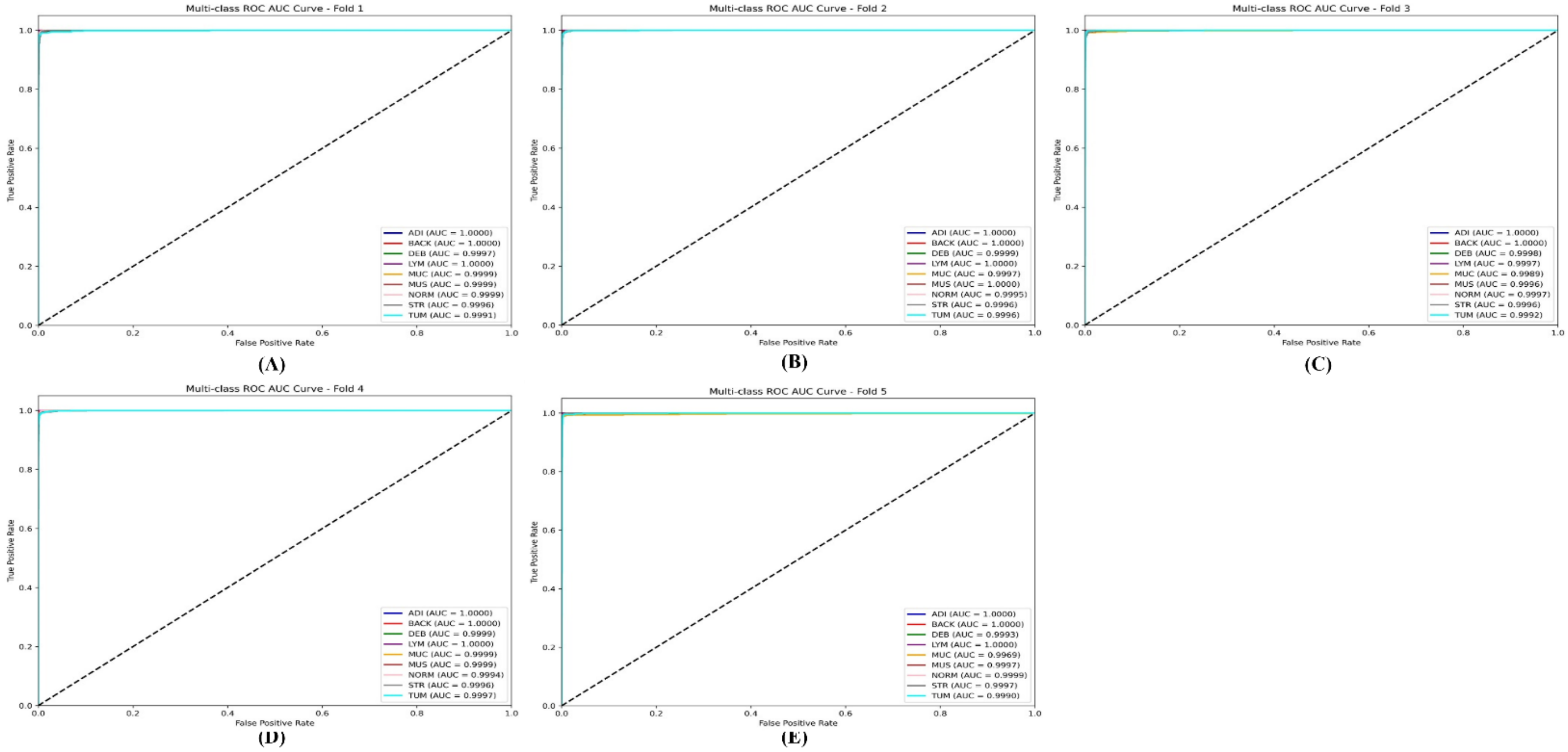

***Figure 11. ROC-AUC curves of MSRANetV2 for nine-class colorectal tissue classification across five folds: (A) Fold 1, (B) Fold 2, (C) Fold 3, (D) Fold 4, and (E) Fold 5 on the NCT-CRC-HE-100K dataset.***

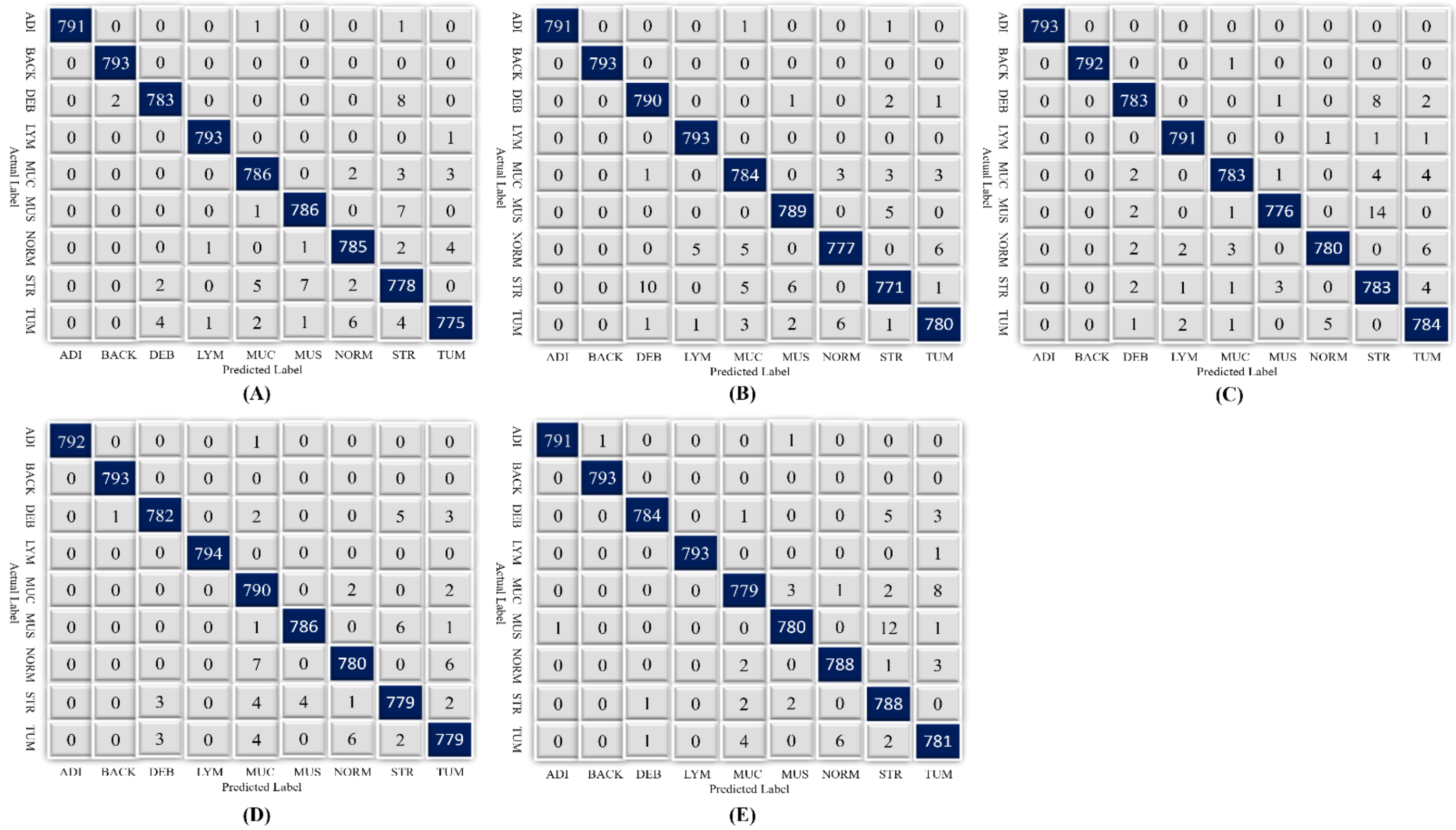

***Figure 12. Confusion matrices showing MSRANetV2's classification performance on five folds: (A) Fold 1, (B) Fold 2, (C) Fold 3, (D) Fold 4, and (E) Fold 5 on the NCT-CRC-HE-100K dataset.***

### 4.3 Explainable AI on MSRANetV2's Interpretability

Gradient-weighted Class Activation Mapping (Grad-CAM) is a technique for visually identifying the portions of an image that a CNN model considers the most relevant when making a classification decision. It works by first running the input image through a CNN to generate a predicted label [25]. The activation maps from the final convolutional layer are weighted and merged to form a heatmap. This heatmap emphasizes the regions with the greatest influence on the prediction and overlays the original image to indicate which parts the model focused on during categorization.

Figure 13 shows Grad-CAM visualizations for colorectal cancer classification using the MSRANetV2 model. Each row represents a separate histopathological class from the nine total classes, of which only three are described here. In each row, the first column displays the original histopathology image, the second column displays the corresponding Grad-CAM heatmap indicating the areas most influential to the model's decision, and the third column is the heatmap overlay on the original image, highlighting MSRANetV2's regions of interest (ROI). The heatmaps have a color gradient from red to blue, with red/yellow representing high activation (regions that contributed the most to the categorization) and blue representing low activation. MSRANetV2, most likely using multi-resolution attention, reliably identifies morphological patterns unique to each cancer subtype. The clarity of the attention maps indicates good spatial localization skills, ensuring that the model is not only correct but also interpretable, which is critical for clinical trust and diagnostic support in colorectal cancer analysis.

On a clinical level, Grad-CAM interpretability enhances the model's feasibility by providing visual confirmation of its decisions. Clinicians can correlate the areas of concern denoted by the heatmap with established diagnostic characteristics (e.g., glandular structures, stromal margins, or inflammatory patterns) to verify whether the model is appropriately attending to relevant histological cues. Such alignment facilitates confidence in artificial intelligence-generated predictions and their adoption as assistive tools within diagnostic workflows. In addition, in cases of uncertainty where pathologists might seek a second opinion, such interpretability allows the model to serve as an explainable reference, as opposed to an opaque "black-box" system. Being able to clarify the justification for a classification—more than simply delivering an output—can decrease diagnostic uncertainty, promote adoption in real-world pathology labs, and facilitate collaborative human–AI decision-making.

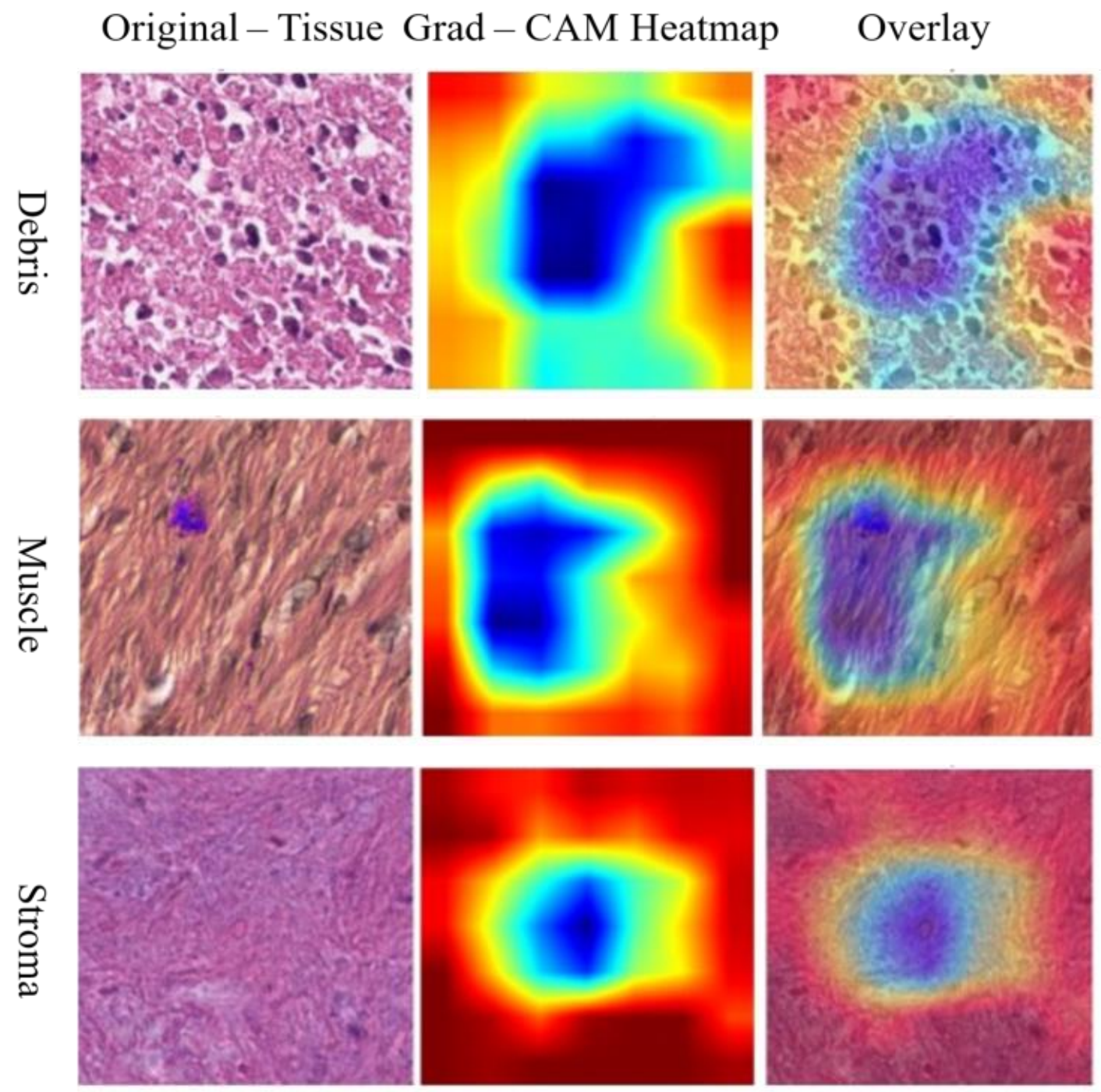

**Figure 13:** ***Grad-CAM visualizations from MSRANetV2 highlight class-discriminatory regions in colorectal histopathology images.***

## 5 Discussions

The discussion section presents an in-depth interpretation of the outcomes obtained from the proposed MSRANetV2 model. It evaluates the model's comprehensive effectiveness in classifying colorectal cancer tissue images and explores how well it generalizes across different dataset variations. Additionally, we reflect on the strengths and constraints of the proposed method. Such insights will help to identify areas for future improvement and research direction.

Apart from its classification accuracy, the proposed MSRANetV2 model is also very computationally efficient. Despite the addition of advanced attention mechanisms and multi-scale fusion strategies, the model still has a modest trainable parameter count of approximately 26.4 million (26,408,201). This is considerably lower than deeper architectures such as ResNet101 or ResNet152, thus rendering MSRANetV2 more computationally viable. Training was efficiently carried out on a high-end GPU (NVIDIA RTX 4090), with no memory bottlenecks, and convergence was typically achieved within 12–15 epochs per fold, with the aid of early stopping and learning rate scheduling. The use of ResNet50V2 as the backbone offers the optimal balance of representational power and resource consumption. All these considerations facilitate MSRANetV2's potential for real-time deployment and potential robustness in resource-constrained clinical environments.

While promising performance has been shown with the suggested MSRANetV2 model on two benchmark datasets, possible biases due to dataset origin and patient populations must be acknowledged. Both the CRC-VAL-HE-7K and NCT-CRC-HE-100K datasets were obtained under specific clinical and institutional settings, and their patient populations might not represent overall ethnic or geographic heterogeneity. Thus, generalization to unseen populations or images from different staining protocols, scanners, or clinical environments might be limited. While five-fold cross-validation has assisted in reducing overfitting and enhancing robustness, future work should consider external validation on independent datasets from multiple sources to further assess generalizability and mitigate hidden biases.

An important aspect to evaluate is the nature and implication of misclassifications revealed through the confusion matrices. Certain classes—such as Stroma (STR), Mucus (MUC), and Tumor Epithelium (TUM)—exhibited occasional confusion with visually or morphologically similar tissues like Muscle (MUS) and Normal Mucosa (NORM). For example, STR was incorrectly classified as MUS or DEB in some folds, most likely as a result of overlapping stromal patterns and a similar fibrous texture. Similarly, MUC samples were confused with TUM or NORM in rare instances, as mucus-producing tumor glands and inflamed mucosa can appear structurally similar under staining. Clinically, these misclassifications can cause diagnostic ambiguity, especially when it comes to mixed or borderline tissue types, where it might be difficult for pathologists to assign labels. Even though these errors were small and the model maintained high overall metrics across folds, enhancing diagnostic trustworthiness requires knowing where the uncertainty is coming from. Integrating additional spatial context, staining variations, or ensemble techniques could potentially mitigate these confusions in future iterations.

### 5.1 Comparative Analysis

To assess the robustness of the proposed MSRANetV2 architecture, an extensive comparative analysis was made with several existing state-of-the-art models. The comparison was made based on important performance measures on different models for histopathology data for colorectal cancer. The proposed MSRANetV2 model was evaluated on both the CRC-VAL-HE-7K and NCT-CRC-HE-100K datasets. It achieved remarkable performance with test accuracy rate of 99.05% on the CRC-VAL-HE-7K dataset and 99.02% on the NCT-CRC-HE-100K dataset. Corresponding precision, recall, and F1-score values were consistently high at $0.9884 \pm 0.0151$, $0.99 \pm 0.0151$, and $0.99 \pm 0.0145$ on the CRC-VAL-HE-7K dataset, and $0.9904 \pm 0.0091$, $0.9900 \pm 0.0071$, and $0.9900 \pm 0.0071$ on the NCT-CRC-HE-100K dataset, respectively. These results surpass other advanced models across all evaluated metrics.

The superior performance of MSRANetV2 is attributed to several key architectural enhancements. Firstly, the residual attention mechanism enables the model to integrate both deep semantic and fine-grained spatial information through the fusion of multi-scale representations between the intermediate layers of ResNet50V2, conv4_block6_out, and conv5_block3_out. The multi-scale strategy is designed to allow the model to maintain high-level context while keeping low-level textures that are critical in histopathological images. Secondly, the application of Squeeze-and-Excitation (SE) blocks allows the network to adaptively recalibrate the feature responses adaptively along the channel dimension, thereby assisting the model in emphasizing the patterns that are medically relevant. Thirdly, the channel alignment via 1×1 convolution and bilinear upsampling guarantees matching spatial dimensions of the features, allowing the feature fusion process to be more effective. Collectively, these design decisions enhance feature discrimination, leading to increased classification accuracy and robustness.

In comparison, Shah et al. [13] evaluated three models—Inception V3, MobileNet, and ResNet50V2—on an 8-class task. The highest accuracy among these was 95.00% using ResNet50V2, along with a precision, recall, and F1-score of 95.00%. While respectable, this performance falls short of our MSRANetV2 on both datasets. Nektarios et al. [26] implemented XGBoost on an 8-class task and reported an accuracy of 89.79%, precision of 89.66%, recall of 89.74%, and an F1-score of 89.64%. These values show a noticeable gap when compared to MSRANetV2's performance, indicating the superior capability of deep learning-based attention mechanisms over traditional machine learning classifiers for histopathological image classification. Elshamy et al. [27] used CNN (SAdagrad) and achieved 98.00% accuracy, 97.00% precision, 98.00% recall, and 98.00% F1-score. Although impressive, our MSRANetV2 model achieved higher or equivalent metrics, affirming its reliability and efficiency in clinical diagnosis settings. Chandradeep et al. [14] explored several architectures, including DenseNet121, Xception, and Inception ResNet V2. DenseNet121 yielded 87.20% accuracy, significantly lower than MSRANetV2. Xception and Inception ResNet V2 offered slightly better results, with 95.20% and 94.20% accuracies respectively. However, even their best F1-scores of 95.00% and 94.00% did not match the 99.00% reported by MSRANetV2. Kumar et al. [28] proposed CRCCN-Net and reported accuracies of 93.50% (8-class) and 96.26% (9-class), with F1-scores peaking at 96.38%. While effective, MSRANetV2 surpassed these metrics, suggesting better multi-class discrimination and generalization capability. Ghosh et al. [29] presented an Ensemble DNN model achieving a 96.16% accuracy, 96.15% precision, and 96.16% F1-score. Though close, MSRANetV2's metrics outperform these across all categories, reflecting its architectural improvements like multi-scale attention and SE blocks. Martinez-Fernandez et al. [30] used VGG-19 and obtained 96.40% accuracy, with an F1-score of 94.44%. Jiang et al. [31] employed GAN with Inception and achieved even lower results—accuracy of 89.54% and F1-score of 88.70%, highlighting the limitations of generative approaches alone for classification. Khalid et al. [32] developed CCDNet and achieved the highest non-MSRA results: 98.96% accuracy and 98.64% F1-score. However, our model still shows a slight edge, especially in recall and balanced F1-score across all classes. Lastly, Dabass et al. [33] with CNN (ECLMS+ALM+TMs) achieved 97.70% accuracy and 97.71% F1-score—still notably below MSRANetV2's 99.00% F1-score.
On the other hand, most state-of-the-art methods exhibited inferior performance, either due to architectural constraints or inefficient attention mechanisms. For example, models such as XGBoost or less deep CNN architectures cannot capture complex hierarchical features, which is a requirement for high-resolution histopathological classification. Those models lacking explicit multi-scale fusion or channel recalibration (e.g., DenseNet121, InceptionResNetV2) fail to distinguish fine structure patterns, particularly in classes with subtle differences such as STR with MUS. Ensemble-based or GAN-inspired approaches introduce additional computational complexity or instability without providing proportional performance gain. This analysis underscores the efficiency of carefully crafted attention-based models, such as MSRANetV2, considering balance among depth, interpretability, and feature specificity.

**Table 5: Evaluation Metrics Comparison Across State-of-the-Art Methods**

| Author | Proposed Model | Number of classes | Evaluation Metrics | | | |
|---|---|---|---|---|---|---|
| | | | Accuracy (%) | Precision (%) | Recall (%) | F1-Score (%) |
| Shah et al. [13] | Inception V3 | 8 | 88.50 | 91.00 | 91.00 | 91.00 |
| | MobileNet | 8 | 91.50 | 91.00 | 91.00 | 91.00 |
| | ResNet50V2 | 8 | 95.00 | 95.00 | 95.00 | 95.00 |
| Nektarios et al. [26] | XGBoost | 8 | 89.79 | 89.66 | 89.74 | 89.64 |
| Elshamy et al. [27] | CNN (SAdagrad) | 8 | 98.00 | 97.00 | 98.00 | 98.00 |
| Chandradeep et al. [14] | DenseNet121 | 9 | 87.2 | 84.70 | 85.60 | 85.00 |
| | Xception | 9 | 95.2 | 95.20 | 95.40 | 95.00 |
| | Inception ResNet V2 | 9 | 94.20 | 92.50 | 95.70 | 94.00 |
| Kumar et al. [28] | CRCCN-Net | 8 | 93.50 | 94.12 | 93.62 | 93.86 |
| | | 9 | 96.26 | 96.44 | 96.34 | 96.38 |
| Ghosh et al. [29] | Ensemble DNN | 8 | 92.83 | 92.83 | 93.11 | 92.83 |
| | | 9 | 96.16 | 96.17 | 96.15 | 96.16 |
| Martínez-Fernandez et al. [30] | VGG-19 | 9 | 96.40 | 94.22 | 94.44 | 94.44 |
| Jiang et al. [31] | InceptionV3-SMSG-GAN | 9 | 89.54 | 86.84 | 86.62 | 98.70 |
| Khalid et al. [32] | CCDNet | 8 | 98.61 | 98.55 | 98.33 | 98.24 |
| | | 9 | 98.96 | **99.37** | 98.80 | 98.64 |
| Dabass et al. [33] | CNN (ECLMS+ALM+TMs) | 9 | 97.70 | 97.69 | 97.73 | 97.71 |
| **Proposed Architecture (Our)** | **MSRANetV2 (7K DS)** | 9 | **99.05** | 98.84 | **99.00** | **99.00** |
| | **MSRANetV2 (100K DS)** | 9 | **99.02** | 99.04 | **99.00** | **99.00** |

### 5.2 Strength & Limitations

The proposed MSRANetV2 model provides an effective mechanism for the classification of colorectal tissue through the combination of multi-scale residual attention, thereby enhancing low-level spatial feature and high-level semantic representation. Through the adoption of feature maps of conv4 and conv5 phases of ResNet50V2, followed by the dimensional alignment and attention recalibration, the model successfully fuses important information to accomplish powerful classification. Its utilization of Squeeze-and-Excitation (SE) blocks enhances performance through channel-wise feature refinement, making the model sensitive to informative regions. Utilizing ImageNet pretrained weights, MSRANetV2 is able to leverage transfer learning, decreasing training time and enhancing convergence. Its end-to-end trainable model simplifies the implementation and makes it scalable to related medical tasks. The model performs very well on both the CRC-VAL-HE-7K and NCT-CRC-HE-100K datasets with high accuracy and F1-score in five folds of stratified cross-validation. The use of global average pooling aids in reducing dimensionality without compromising on critical features necessary for the final classification. The model demonstrates robustness against class imbalance and fold consistency across various folds. The use of explainable AI through Grad-CAM also allows for visualizing the decision-making of the model. Through careful hyperparameters selection and minimal preprocessing, MSRANetV2 attains state-of-the-art performance at the cost of computational efficiency. Its modularity makes easy adaption of the attention mechanisms, permitting flexibility for further research. The model's robustness, scalability, and higher classification metrics make it a reliable choice for histopathological image analysis. Its compatibility in standard computing setups also makes it increasingly accessible to research and clinical applications. Overall, MSRANetV2 achieves an excellent balance among accuracy, interpretability, and architectural novelty.

As much as its promising performance, MSRANetV2 is not without limitations. The ResNet50V2 backbone, albeit powerful, increases the computational overhead, thereby limiting its real-time application on resource-constrained devices. Although Grad-CAM visualizations are informative, they nevertheless require manual adjustment for clinically meaningful interpretation. Lastly, optimization of inference speed is a future task in the case of clinical deployment.

The generalizability of the MSRANetV2 across different clinical contexts is another drawback. On images from various staining techniques, scanners, or patient demographics, its performance could deteriorate. For multi-institutional, ethnically diverse datasets to have wider relevance, external validation is necessary.

The incorporation of deeper feature extractors and attention mechanisms in subsequent expansions may result in an increase in MSRANetV2's computing cost. Nonetheless, methods like pruning and model quantization can be used to reduce memory and inference time. These optimizations can facilitate deployment in real-time or resource-constrained clinical environments without significantly sacrificing accuracy.

## 6 Conclusion

This study proposed a robust MSRANetV2 architecture for the accurate classification of colorectal cancer tissues using histopathological images. By integrating residual attention mechanisms, multi-scale feature extraction from ResNet50V2, and Squeeze-and-Excitation (SE) blocks, the model effectively captures both low-level textures and high-level semantic features. Experimental results on two benchmark datasets CRC-VAL-HE-7K and NCT-CRC-HE-100K demonstrated exceptional classification performance, with average F1-scores and AUC values exceeding 0.99 across all folds. The model not only outperformed conventional CNN-based architectures but also exhibited robustness against class imbalance and noise, underscoring its practical applicability in clinical pathology. Furthermore, the incorporation of Grad-CAM-based visual explanations enhanced the interpretability of predictions, offering transparency crucial for medical decision support. Future research can extend this work by exploring transformer-based modules and domain adaptation techniques to generalize across diverse histopathological datasets and institutions. In essence, this approach significantly advances digital pathology, offering a step toward more accurate, interpretable, and accessible AI-driven cancer diagnostics.

**Declaration of interests:** The authors declare that they have no known competing financial interests or personal relationships that could have appeared to influence the work reported in this paper.

**Data Availability:** Data will be made available on request.